\PassOptionsToPackage{kerning}{microtype}

\documentclass[conference]{IEEEtran}

\usepackage{amsmath,amssymb,amsfonts}
\usepackage{algorithmic}
\usepackage{graphicx}
\usepackage{textcomp}
\usepackage{xcolor}
\def\BibTeX{{\rm B\kern-.05em{\sc i\kern-.025em b}\kern-.08em
T\kern-.1667em\lower.7ex\hbox{E}\kern-.125emX}}

\usepackage[main=UKenglish]{babel}
\usepackage{xspace}
\usepackage[%
  group-minimum-digits=4,%
  list-final-separator={, and },%
  print-zero-integer=true,%
  free-standing-units,%
  round-mode=figures,%
  round-precision=3,%
  mode=match,%
  reset-text-series=false,%
  text-series-to-math=true,%
  uncertainty-mode=separate,%
  round-pad=false,%
]{siunitx}
\usepackage[]{subcaption}
\usepackage{listings}
\lstdefinestyle{inlinelst}{%
  basicstyle=\normalsize\ttfamily,%
  keywordstyle=,%
  language=Python,%
}
\lstdefinestyle{numberedlst}{%
  language=Python,%
  basicstyle=\scriptsize\ttfamily,%
  numbers=left,%
  numberstyle=\tiny,%
  tabsize=2,%
  breaklines=true,%
  xleftmargin=1.9em,%
  xrightmargin=1em,%
}
\lstdefinestyle{nonnumberedlst}{%
  language=Python,%
  basicstyle=\scriptsize\ttfamily,%
  numbers=none,%
  tabsize=2,%
  breaklines=true,%
  xleftmargin=1em,%
  xrightmargin=1em,%
}
\lstdefinestyle{nonnumberedtxt}{%
  basicstyle=\scriptsize\ttfamily,%
  numbers=none,%
  tabsize=2,%
  breaklines=true,%
  xleftmargin=1em,%
  xrightmargin=1em,%
}
\NewDocumentCommand{\inlinelst}{m}{\lstinline[style=inlinelst]{#1}}
\usepackage{paralist}
\usepackage{balance}
\usepackage{colortbl}
\usepackage{adjustbox}
\usepackage{multirow}
\usepackage{todonotes}
\usepackage{csquotes}
\usepackage{mismath}
\usepackage{microtype}
\SetExtraKerning{
  encoding = {T1}
}{
  \textemdash = {167,167},
  — = {167,167}
}
\usepackage{tikz}
\usetikzlibrary{arrows.meta, positioning, shapes.multipart,
shapes.geometric, calc, backgrounds}
\usepackage{booktabs}

\usepackage{enumitem}
\newlist{rqlist}{enumerate}{1}
\setlist[rqlist,1]{label=\textbf{RQ\arabic*:}, ref=RQ\arabic*, nosep, left=0pt}

\let\origfootnote\footnote
\renewcommand{\footnote}[1]{\kern.06em\origfootnote{#1}}
\newcommand{\punctfootnote}[1]{\kern-.20em\origfootnote{#1}}

\usepackage{tcolorbox}
\usepackage{xspace}
\usepackage{relsize}

\newcommand{\toolname}[1]{%
  \textsc{#1}\xspace%
}

\newcommand{\ffiname}[1]{%
  \texttt{#1}\xspace%
}

\newcommand{\pynguin}{%
  \toolname{Pynguin}%
}

\newcommand{\Pynguin}{%
  \toolname{Pynguin}%
}

\newcommand{\tensorflow}{%
  \toolname{TensorFlow}%
}

\newcommand{\pyTorch}{%
  \toolname{PyTorch}%
}

\newcommand{\evosuite}{%
  \toolname{EvoSuite}%
}

\newcommand{\randoop}{%
  \toolname{Randoop}
}

\newcommand{\numpy}{%
  \toolname{NumPy}%
}

\newcommand{\scipy}{%
  \toolname{SciPy}%
}

\newcommand{\pandas}{%
  \toolname{Pandas}%
}

\newcommand{\codamosa}{%
  \toolname{CodaMosa}
}

\newcommand{\dsc}{%
  \textsc{DS-C}\xspace
}

\newcommand{\dscodamosa}{%
  \textsc{DS-Co\-da\-Mo\-sa}\xspace
}

\newcommand{\subprocessexec}{%
  sub\-proc\-ess-exe\-cu\-tion\xspace
}

\newcommand{\Subprocessexec}{%
  Sub\-pro\-cess-exe\-cu\-tion\xspace
}

\newcommand{\SubprocessExec}{%
  Sub\-pro\-cess-Exe\-cu\-tion\xspace
}

\newcommand{\subprocess}{%
  sub\-proc\-ess\xspace
}

\newcommand{\Subprocess}{%
  Sub\-proc\-ess\xspace
}

\newcommand{\threadedexec}{%
  threa\-ded-exe\-cu\-tion\xspace
}

\newcommand{\threaded}{%
  threa\-ded\xspace
}

\newcommand{\heuristic}{%
  heuristic\xspace
}

\newcommand{\Heuristic}{%
  Heuristic\xspace
}

\newcommand{\fallback}{%
  re\-start\xspace
}

\newcommand{\Fallback}{%
  Re\-start\xspace
}

\newcommand{\fallbackheuristic}{%
  combined\xspace
}

\newcommand{\Fallbackheuristic}{%
  Combined\xspace
}

\newcommand{\crashrev}{%
  cra\-sh-re\-vea\-ling\xspace
}

\newcommand{\CrashRev}{%
  Cra\-sh-Re\-vea\-ling\xspace
}

\newcommand{\CExtension}{%
  C-ex\-ten\-sion\xspace
}

\newcommand{\CExtensions}{%
  C-ex\-ten\-sions\xspace
}

\newcommand{\pypi}{%
  \toolname{PyPI}%
}

\newcommand{\pytest}{%
  \toolname{pytest}%
}

\newcommand{\SUT}{SUT\xspace}
\newcommand{\SUTs}{SUTs\xspace}
\newcommand{\FFI}{FFI\xspace}
\newcommand{\FFIs}{FFIs\xspace}

\newcommand\effectsize{\ensuremath{\hat{A}_{12}}\xspace}

\newcommand*{\eg}{e.g.\xspace}
\newcommand*{\ie}{i.e.\xspace}

\NewDocumentEnvironment{summary}{m}{%
  \begin{tcolorbox}[title={Summary~(#1)}]%
  }{%
  \end{tcolorbox}%
}

\newcommand\CPP{%
  C\nolinebreak[4]\hspace{-.05em}%
  \raisebox{.4ex}{\relsize{-3}{\textbf{++}}}\xspace%
}

\newcommand{\fault}[1]{\textit{#1}\xspace}

\newcommand{\segfault}{\fault{segmentation fault}}
\newcommand{\segfaults}{\fault{segmentation faults}}

\newcommand{\Segfaults}{\fault{Segmentation faults}}

\newcommand{\SegFaults}{\fault{Segmentation Faults}}

\newcommand{\aborteds}{\fault{aborteds}}

\newcommand{\illegalinstructions}{\fault{illegal instructions}}

\newcommand{\buserrors}{\fault{bus errors}}

\newcommand{\fpes}{\fault{floating-point exceptions}}

\newcommand{\oomkills}{\fault{OOM kills}}
\newcommand{\longoomkills}{\fault{out of memory (OOM) kills}}

\newcommand{\timeouts}{\fault{timeouts}}

\newcommand{\memleaks}{\fault{memory leaks}}

\newcommand{\assertionfailure}{\fault{native assertion failures}}

\newcommand{\exnum}[1]{\num[round-mode=none]{#1}}

\newcommand{\modules}{\exnum{1648}\xspace}
\newcommand{\codamosaModules}{486}
\newcommand{\removedModules}{\exnum{1005}\xspace} %
\newcommand{\libraries}{\exnum{21}\xspace}
\newcommand{\pypiLibsInitial}{\exnum{500}\xspace}
\newcommand{\identifiedFaults}{\exnum{32}\xspace}

\newcommand{\CTrivialModules}{\exnum{95}\xspace}
\newcommand{\CExecutionTime}{41268}
\newcommand{\CodamosaExecutionTime}{12170}
\newcommand{\CDefaultSuccess}{\exnum{625}\xspace}
\newcommand{\CSubprocessSuccess}{\exnum{1438}\xspace}
\newcommand{\CCrashesReducedPercent}{\qty{56.53685674547984}{\percent}\xspace}
\newcommand{\CModulesOnlyInSubprocess}{\exnum{215}\xspace}

\newcommand{\CSubprocessBetterSig}{\exnum{987}\xspace}
\newcommand{\CSubprocessBetter}{\exnum{1072}\xspace}

\newcommand{\CSubprocessWorse}{\exnum{331}\xspace}
\newcommand{\CSubprocessWorseSig}{\exnum{247}\xspace}
\newcommand{\CSubprocessEffectSize}{\num{0.6934688511326853}\xspace}

\newcommand{\CodamosaSubprocessBetterSig}{\exnum{37}\xspace}
\newcommand{\CodamosaSubprocessBetter}{\exnum{53}\xspace}

\newcommand{\CodamosaSubprocessWorse}{\exnum{199}\xspace}
\newcommand{\CodamosaSubprocessWorseSig}{\exnum{161}\xspace}
\newcommand{\CodamosaSubprocessEffectSize}{\num{0.418207590306356}\xspace}

\newcommand{\CHeuristicSubprocessPercent}{\qty{73.94619741100324}{\percent}\xspace}
\newcommand{\CodamosaHeuristicSubprocessPercent}{\qty{63.26474622770919}{\percent}\xspace}

\newcommand{\CRestartTriggeredModulesPercent}{\qty{59.101941747572816}{\percent}\xspace}
\newcommand{\CodamosaRestartTriggeredModulesPercent}{\qty{11.11111111111111}{\percent}\xspace}

\newcommand{\CrashRevealingTotalSubprocess}{\exnum{120176}\xspace}

\newcommand{\CrashRevealingNonReproducibleSubprocess}{\exnum{5465}\xspace}

\newcommand{\CrashRevealingReproducibleCountSubprocess}{\exnum{114711}\xspace}

\newcommand{\CrashRevealingCauseCount}{\exnum{213}\xspace}

\newcommand{\AbortedCount}{\exnum{25}\xspace}

\newcommand{\SegmentationFaultCount}{\exnum{185}\xspace}
\newcommand{\TimeoutCount}{\exnum{3}\xspace}

\newcommand{\AbortedPercent}{\qty{11.737089201877934}{\percent}\xspace}

\newcommand{\SegmentationFaultPercent}{\qty{86.85446009389672}{\percent}\xspace}
\newcommand{\TimeoutPercent}{\qty{1.4084507042253522}{\percent}\xspace}

\usepackage{hyperref}
\usepackage[capitalise]{cleveref}

\IEEEoverridecommandlockouts

\begin{document}

\title{Real-World Fault Detection for C-Extended Python Projects with
Automated Unit Test Generation}

\author{
  \IEEEauthorblockN{
    Lucas Berg\IEEEauthorrefmark{1}\(^{\perp}\),
    Lukas Krodinger\IEEEauthorrefmark{2}\(^{\perp}\),
    Stephan Lukasczyk\IEEEauthorrefmark{3},
    Annibale Panichella\IEEEauthorrefmark{4},\\
    Gordon Fraser\IEEEauthorrefmark{2},
    Wim Vanhoof\IEEEauthorrefmark{1} and
    Xavier Devroey\IEEEauthorrefmark{1}
  }
  \IEEEauthorblockA{
    \IEEEauthorrefmark{1}NADI, University of Namur,
    Namur, Belgium
  }
  \IEEEauthorblockA{
    \IEEEauthorrefmark{2}University of Passau,
    Passau, Germany
  }
  \IEEEauthorblockA{
    \IEEEauthorrefmark{3}JetBrains Research,
    Munich, Germany
  }
  \IEEEauthorblockA{
    \IEEEauthorrefmark{4}Delft University of Technology,
    Delft, Netherlands
  }
  \thanks{\(^{\perp}\)The first two authors contributed equally to this
  work.} %
}

\maketitle

\begin{abstract}
  Many popular Python libraries use \CExtensions for performance-critical
  operations allowing users to combine the best of the two worlds:
  The simplicity and versatility of Python and the performance of C\@.
  A drawback of this approach is that exceptions raised in C can
  bypass Python's exception handling and cause the entire interpreter to crash.
  These crashes are real faults if they occur when calling a public API\@.
  While automated test generation should, in principle, detect such faults,
  crashes in native code can halt the test process entirely, preventing
  detection or reproduction of the underlying errors and inhibiting coverage
  of non-crashing parts of the code.
  To overcome this problem, we propose separating the generation and
  execution stages of the test-generation process.
  We therefore adapt \pynguin, an
  automated test case generation tool for Python, to use \subprocessexec.
  Executing each generated test in an isolated subprocess prevents a
  crash from halting the test generation process itself. This allows us
  to (1) detect such faults, (2) generate reproducible \crashrev test
  cases for them, (3) allow studying the underlying faults, and
  (4) enable test generation for non-crashing parts of the code.
  To evaluate our approach, we created a dataset consisting of \modules modules
  from \libraries popular Python libraries with C-extensions.
  \Subprocessexec allowed automated testing of up to \CCrashesReducedPercent
  more modules and discovered \CrashRevealingCauseCount unique crash
  causes, revealing \identifiedFaults
  previously unknown faults. 
\end{abstract}


\section{Introduction}\label{sec:introduction}

Python has emerged as one of the most widely used programming languages
in scientific
computing~\cite{stackoverflow_dev_survey_2023},
data analysis, and artificial intelligence, primarily
due to its simplicity and extensive ecosystem of libraries. However, as an
interpreted language, Python may not always provide the performance
necessary for computationally intensive tasks. To mitigate these
performance bottlenecks, Python's \emph{foreign function
interface}~(\FFI) allows it to interface with
compiled languages, particularly C/\CPP.
Many Python libraries, especially in the
scientific computing and machine-learning domains, like
\numpy~\cite{numpy},
\pandas~\cite{pandas},
and \tensorflow~\cite{tensorflow}
use \FFIs.

However, the \FFI can lead to a crash of the
Python interpreter if misused. For example, \Cref{fig:ffi-example}
shows the function \inlinelst{idd_reconid} from the scientific computing
library \scipy~\cite{scipy}.
It is written in \ffiname{Cython}~\cite{cython_tool}, a superset of Python
that compiles to C and allows developers to write C-extension modules
using a Python-like syntax. After compilation, developers can import
\inlinelst{idd_reconid} from its C-extension module and call it from
Python code as if it were a regular Python function.
This ability of the Python interpreter to import and invoke functions
written in other languages is made possible by Python's \FFI.
However, the function \inlinelst{idd_reconid} does not validate its
first parameter: passing a one-dimensional array instead of a \numpy
matrix can result in a \segfault at line~4.
The crash occurs because the Python interpreter executes native code,
thereby bypassing Python's built-in memory-safety mechanisms.
Unlike Python, which provides automatic memory management and bounds
checking, C requires manual memory allocation and permits direct memory access.
In this example, \scipy enables a Cython optimisation that disables bounds
checking to improve performance. As a result, indexing operations
follow C semantics,
causing \segfaults instead of raising
\inlinelst{IndexErrors}~\cite{cython_compiler_directives}.
This example illustrates only one type of fault that can arise
from misusing Python's \FFI; in addition to \segfaults, other low-level errors
such as \buserrors, \memleaks, and \fpes can also occur.
All these faults are particularly severe because they can
cause the Python interpreter to crash, which poses a threat
to software reliability.

\begin{figure}[t]
  \centering
\begin{lstlisting}[
    style=numberedlst,%
    frame=tb,%
  ]
# cython: boundscheck=False

def idd_reconid(B, idx, proj):
    cdef int m = B.shape[0], krank = B.shape[1]
    cdef int n = len(idx)
    approx = np.zeros([m, n], dtype=np.float64)

    approx[:, idx[:krank]] = B
    approx[:, idx[krank:]] = B @ proj

    return approx
\end{lstlisting}
  \caption{The \inlinelst{idd_reconid} function from \scipy, which is
    included in a C-extension module, can cause a
    \segfault at line 4 if the argument \texttt{B} is not a valid \numpy matrix.
  }
  \label{fig:ffi-example}
\end{figure}

To mitigate this risk, diagnosis tools were developed, especially
to detect \memleaks~\cite{mitchell_leakbot_2003},
using binary instrumentation~\cite{nethercote_valgrind_2007},
static analysis~\cite{hu_pythonc_2020},
and dynamic analysis for Java~\cite{yu_dynamic_2021}
and Python~\cite{memray_github}.
However, these tools do not automatically find issues and require manual
intervention: developers must already have a failing test or know how to
reproduce an issue to apply a memory-leak diagnosis tool.
This leaves developers to manually locate
faults and reverse-engineer the conditions that trigger them,
which is time-consuming and error-prone.

To alleviate developers' workload and manage effort and costs,
automated unit test generation tools have been introduced.
Notable examples of existing tools include \randoop~\cite{Pacheco2007} and
\toolname{EvoSuite}~\cite{fraser_evosuite_2011} for
Java or \pynguin~\cite{lukasczyk_pynguin_2022} for Python. These tools
typically examine a \emph{subject under test}~(\SUT), such as a Java
class or Python module, and use meta-heuristic search algorithms to
generate test cases. The objective is to create inputs that trigger
the \SUT's routines (i.e., functions, methods, and constructors) in a
way that optimises a specific fitness metric, most commonly branch coverage.
During coverage optimisation, the tools iteratively generate, modify,
and execute tests while measuring their coverage.
Consequently, gradually more parts of the \SUT are tested.
Exceptions triggered during execution may indicate faults in the SUT\@,
which holds for search- and LLM-based test generation approaches and
was previously explored for Java~\cite{fraser_evosuite_2013,
fraser_1600_2015}, but not yet for Python.%

\Pynguin~\cite{lukasczyk_pynguin_2022} generates tests for Python
and, like any other Python-based tool, requires an interpreter to run
and execute generated tests.
However, a \segfault
or a similar C error uncaught by Python's
exception handling
can crash the Python interpreter, which in turn terminates the
test-generation process and thus prevents the detection of the underlying fault
causing the crash.
In Java, where \FFIs are less common, tools like \evosuite
did not require process isolation in past studies. In contrast,
Python's extensive
use of \FFIs means that
interpreter crashes due to C errors pose a significant challenge for
automated test generation tools like \pynguin, hindering developers to
detect, quantify and investigate bugs in C-extended projects.

By executing the \SUT in a separate subprocess rather than only in a separate
thread, we isolate the test generation process from the \SUT's execution
and thereby overcome this limitation of \pynguin.
The architectural separation acts as a sandbox,
protecting the test generator itself. If a test case triggers a fatal
error that cannot be caught from its Python invocation point, like a
\segfault in the \SUT, only the subprocess
terminates, leaving the main search process unharmed. Our approach
monitors these subprocesses and detects crashes. Upon detection,
the corresponding test case that triggered the fault is exported as a
\toolname{pytest}~\cite{pytest}
test case. Re-execution ensures that the test case is reproducible, so
developers can use it to debug and fix underlying issues.
Overall, our approach has four advantages compared to running \pynguin
without \subprocessexec :
(1)~it allows \pynguin to detect \crashrev faults in the \SUT,
(2)~it enables the generation of \crashrev test cases,
(3)~it enables the study of the underlying faults, and
(4)~it makes it possible to continue the test-generation process
even if the \SUT crashes.
These advantages come with a drawback: \subprocessexec
requires more execution time than \threadedexec due to
process communication overhead.
We examine when subprocess execution is beneficial and present
multiple strategies to automate the decision of which execution
strategy to use.
While spawning processes is not difficult, and done in
fuzzing~\cite{zhu_fuzzing_2022}, it is
challenging in the context of test generation because of the code
instrumentation for fitness calculations.
This, however, is crucial to allow \pynguin to explore the \SUT
without the need to manually create data providers, as in fuzzing.
Previous studies on
\pynguin~\cite{lukasczyk_automated_2020,lukasczyk_pynguin_2022,lukasczyk_empirical_2023,lemieux_codamosa_2023}
did not focus on Python modules that use the Python \FFI.
We address this gap by explicitly targeting such modules.
Although our work is grounded in the Python ecosystem, it tackles the broader
challenge of robustly testing software that integrates native code via an \FFI.

To evaluate our approach, we executed \pynguin with and without \subprocessexec
on a large dataset of \modules Python modules from \libraries popular
libraries using C-extensions.
We created this dataset as we are not aware of any existing datasets of
popular real-world
Python projects with C-extensions.
Using \subprocessexec, we automatically generated
\CrashRevealingTotalSubprocess tests revealing
\CrashRevealingCauseCount crash causes. We
manually analysed these to identify \identifiedFaults
previously unknown faults. For
example, \crashrev tests
raising \segfaults allowed us to discover
that the \inlinelst{idd_reconid} function of \scipy, shown in
\cref{fig:ffi-example}, does not check its parameters properly.
To validate our findings, we reported these bugs to the respective
development teams.

In detail, the contributions of this paper are:
\begin{compactitem}
\item We introduce a new \subprocessexec for \pynguin that allows
  generating test cases that can reveal faults.
\item We create a new dataset of \modules Python modules from popular
  libraries that use C-extensions.
\item We conduct a large-scale study of the faults detected by
  using \pynguin with \subprocessexec on the dataset.
\item We identify \identifiedFaults previously unknown faults in
  the libraries using the generated test cases.
\end{compactitem}
Our evaluation demonstrates that the \subprocessexec of \pynguin is
effective in automatically revealing faults in Python libraries that
use C-extensions.
We provide additional artefacts consisting of datasets, tool images,
result data, and analysis scripts to Zenodo to allow for future
use~\cite{dataset}.

\section{Background}\label{sec:background}

Our approach is based on unit test generation for Python programs and
explicitly tailored to Python modules using the Python \emph{foreign function
interface}~(\FFI) to execute C/\CPP code.

\subsection{Foreign Function Interfaces}\label{sec:background:ffi}

A \emph{foreign function interface}~(\FFI) is a mechanism that allows
code written in one programming language to call functions and access
data structures written in another.
The concept is not specific to Python~\cite{jni_docs, jna_docs,
haskell_ffi}, but it is particularly relevant to this language, as it
allowed its ecosystem to expand into areas that usually require
fast programming languages, such as artificial intelligence and data
analysis. By interfacing with C/\CPP code, many Python libraries offer
a user-friendly and expressive interface while
still delivering high performance. This is mainly what makes
Python the most widely used programming language in
scientific computing~\cite{stackoverflow_dev_survey_2023}.

There are several mechanisms for creating \FFI bindings in Python,
each with its trade-offs in terms of ease of use, performance,
and portability~\cite{cython_tool, python_ctypes, cffi_docs,
python_extension_modules, swig_tool}. However, their use introduces
challenges as they rely on native code. For instance, while
Python's exception system is designed to handle runtime errors via
\inlinelst{try}-\inlinelst{except} blocks, it cannot intercept
failures that originate in native code such as \segfaults. While the
Python \ffiname{faulthandler}~\cite{python_faulthandler} module can
dump the Python stack trace when such faults occur, it does not
prevent the interpreter from crashing.
Importantly, \segfaults are merely one example from a
broad spectrum of native faults. Other fatal issues include
\memleaks, \illegalinstructions, \buserrors, and \assertionfailure.
These faults are particularly critical because they undermine the
reliability of any Python application and can lead to security
vulnerabilities such as remote code execution. It is therefore
essential to rigorously test code that uses the \FFI bindings.

\subsection{Unit Test Generation for
Python}\label{sec:background:testgeneration}

Creating strong test suites is essential for minimising the risk of
software failures. However, the process of writing test cases manually
can be labour-intensive. To address this, various automated
approaches for generating unit tests have been developed~\cite{FR19}.
Some methods rely on random generation, potentially augmented by
feedback from previous executions to guide future test
creation~\cite{PLE+07}. Other techniques employ \emph{search-based software
engineering}~(SBSE) techniques, such as evolutionary
algorithms, which seek to optimise test cases according to a defined
fitness function~\cite{Ton04}, such as branch coverage.
For an in-depth discussion of search-based test generation, we refer
to the extensive literature on this topic~\cite{CGA18,
Panichella2018}.
In recent years, research has explored
the usage of generative models~\cite{DDS21} and \emph{large language
models}~(LLMs) for test generation~\cite{dakhel_effective_2024,
  yang_enhancing_2024, pizzorno_coverup_2024, lemieux_codamosa_2023,
  ryan_code-aware_2024, yang_llm-enhanced_2025, xiao_optimizing_2024,
  jain_testforge_2025, sapozhnikov_testspark_2024, abdullin_test_2025,
  chen_chatunitest_2024, ouedraogo_llms_2024, roychowdhury_static_2025,
schafer_empirical_2024}.
However, regardless of the used approach, the generated tests must
be run against the \SUT to measure their effectiveness, and therefore
all approaches can be affected by potential crashes due to \FFIs misuse.
Previous
studies~\cite{lukasczyk_automated_2020,lukasczyk_pynguin_2022,lukasczyk_empirical_2023,lemieux_codamosa_2023}
did not consider code that uses \FFI bindings.

\Pynguin~\cite{lukasczyk_pynguin_2022} is a hybrid state-of-the-art
test-generation framework specifically designed for the Python
programming language.  It incorporates multiple SBSE test-generation
strategies, including feedback-directed random test
generation~\cite{PLE+07} and the DynaMOSA~\cite{PKT18b} evolutionary
algorithm, as well as LLM-based and hybrid test-generation strategies
such as CodaMosa~\cite{lemieux_codamosa_2023}.  A fundamental
requirement for all strategies is the ability to execute generated
tests against the \SUT.  During this execution,
\pynguin measures the code coverage achieved by the generated test
cases and uses the feedback on their effectiveness to further improve
them, aiming to produce test cases with high coverage.
However, \pynguin faces challenges when the execution of a \SUT
triggers a C error, such as a \segfault. Such errors cause the Python
interpreter, which also runs \pynguin itself, to crash.

\section{\CrashRev Test Generation}
\label{sec:approach}

In situations where a \SUT could cause the Python interpreter to
crash, the effectiveness of \pynguin is limited. We developed
\subprocessexec to overcome this. It improves \pynguin's robustness
and allows it to create \crashrev tests.

\subsection{Test-Execution Model}
\label{sec:approach:execution}

As the \subprocessexec is based on \pynguin's existing
\threadedexec, we first describe this original execution model.
The test generation is an iterative feedback loop, orchestrated
by an observer system responsible for tracking and collecting
diverse information, such as coverage, executed bytecode, or
assertion results during test execution.
\Cref{fig:threaded-execution} shows the original execution model of
\pynguin.
At the start of the test generation, \pynguin
initialises a set of observers~(1) that monitor the test-generation
process and collect relevant information.
For each test case execution \pynguin \emph{starts} a
\emph{Test Executor Thread}~(2), which executes the test case~(3) in a
separate thread to enable control of the test-case execution time.
Additionally, observers responsible for collecting test-execution results
are handed over.
During the execution of the test case, the \SUT is monitored through
on-the-fly bytecode instrumentation checking for achieved
fitness.
The instrumentation injects hooks into the subject's code (not shown
in the figure), which allows
observers to collect relevant execution
data, including fitness values, executed bytecode, and
assertion outcomes (4).
These data are then combined into \emph{results} and
sent back to the main thread (5), where they are used to guide the
test-generation process. Observers track metadata for the
stopping conditions, such as the current iteration of the genetic
algorithm, and some general data that must persist between
iterations, such as inferred return types (6).

\subsection{\SubprocessExec}
\label{sec:approach:subprocess}

\begin{figure}[t]
\begin{subcaptionblock}{\linewidth}
  \centering
  \scalebox{0.65}{\begin{tikzpicture}[
    process/.style={draw, fill=white, rounded corners, text
    centered, inner sep=6pt},
    arrow/.style={-{Latex}, thick},
    textnode/.style={fill=white, inner sep=1pt}
  ]

  \node[process] (main) {Main Process};
  \node[process, below right=0.8cm and 1.2cm of main, align=center]
  (thread) {Test Executor\\Thread};

  \node[process, below left=0.3cm and 1.1cm of main, align=center]
  (observer_back) {Observers};
  \node[process, below left=0.2cm and 1.0cm of main, align=center]
  (observer) {Observers};
  \node[process, below left=0.1cm and 0.9cm of main, align=center]
  (observer_front) {Observers};

  \path (main.south |- observer_front) coordinate (startCreateObserver);
  \path (observer_front.east) coordinate (endCreateObserver);
  \draw[arrow] (startCreateObserver) -- (endCreateObserver)
    node[textnode, midway, above] {(1) create()};

  \path (observer.south) ++(0.0, -0.4) coordinate (startSendObserver);
  \path (startSendObserver -| main.south) coordinate (endSendObserver);
  \draw[arrow,dashed] (startSendObserver) -- (endSendObserver)
    node[textnode, midway, above] {};

  \path (main.south |- thread.west) coordinate (startCreateThread);
  \path (thread.west) coordinate (endCreateThread);
  \draw[arrow] (startCreateThread) -- (endCreateThread)
    node[textnode, midway, above] {(2) start(test)};

  \path (startCreateThread) ++(0.0, -0.5) coordinate (startSendTestCase);
  \path (startSendTestCase -| thread.south) coordinate (endSendTestCase);

  \path (endSendTestCase) ++(0.0, -0.6) coordinate (executeTest);
  \node[textnode] at (executeTest) {(3) execute(test)};

  \path (executeTest) ++(0.0, -0.4) coordinate (startCallTObservers);
  \path (startCallTObservers -| observer) coordinate (endCallTObservers);
  \draw[arrow] (startCallTObservers) -- (endCallTObservers)
    node[textnode, midway, above] {(4) notify(test)};

  \path (endCallTObservers) ++(0.0, -0.2) coordinate (startRCallTObservers);
  \path (startRCallTObservers -| executeTest) coordinate (endRCallTObservers);
  \draw[arrow,dashed] (startRCallTObservers) -- (endRCallTObservers)
    node[textnode, midway, above] {};

  \path (endRCallTObservers) ++(0.0, -0.6) coordinate (startResultsToMain);
  \path (startResultsToMain -| main.south) coordinate (endResultsToMain);
  \draw[arrow,dashed] (startResultsToMain) -- (endResultsToMain)
    node[textnode, midway, above] {(5) return results};

  \path (endResultsToMain) ++(0.0, -0.1) coordinate (startCallObservers);
  \path (startCallObservers -| observer) coordinate (endCallObservers);
  \draw[arrow] (startCallObservers) -- (endCallObservers)
    node[textnode, midway, above] {(6) notify(results)};

  \path (main.south) ++(0,-4) coordinate (lifelineEndMain);
  \begin{scope}[on background layer]
    \draw[dashed] (main.south) -- (lifelineEndMain);
    \draw[dashed] (thread.south) -- (startResultsToMain);
  \end{scope}

  \draw[dashed] (observer_back.south) -- (lifelineEndMain -| observer_back.south);
  \draw[dashed] (observer.south) -- (lifelineEndMain -| observer.south);
  \draw[dashed] (observer_front.south) -- (lifelineEndMain -| observer_front.south);

\end{tikzpicture}}
  \caption{\label{fig:threaded-execution}%
    \Pynguin's \threadedexec model%
  }
\end{subcaptionblock}
\begin{subcaptionblock}{\linewidth}
  \centering
  \scalebox{0.65}{  \begin{tikzpicture}[
      process/.style={draw, fill=white, rounded corners, text centered,
      inner sep=6pt},
      arrow/.style={-{Latex}, thick},
      textnode/.style={fill=white, inner sep=1pt}
    ]

    \node[process] (main) {Main Process};
    \node[process, below right=3.2cm and 0.7cm of main, align=center]
    (subprocess) {Test Executor\\Subprocess};
    \node[process, below right=1.5cm and 0.7cm of subprocess,
    align=center] (thread) {Test Executor\\Thread};

    \node[process, below left=0.8cm and 0.7cm of main, align=center]
    (observer_back) {Observers};
    \node[process, below left=0.7cm and 0.6cm of main, align=center]
    (observer) {Observers};
    \node[process, below left=0.6cm and 0.5cm of main, align=center]
    (observer_front) {Observers};

    \node[process, below left=0.8cm and 0.85cm of observer, align=center]
    (robserver_back) {Remote \\ Observers};
    \node[process, below left=0.7cm and 0.75cm of observer, align=center]
    (robserver) {Remote \\ Observers};
    \node[process, below left=0.6cm and 0.65cm of observer, align=center]
    (robserver_front) {Remote \\ Observers};

    \path (main.south |- observer_front) coordinate (startCreateObserver);
    \path (observer_front.east) coordinate (endCreateObserver);
    \draw[arrow] (startCreateObserver) -- (endCreateObserver)
      node[textnode, midway, above] {\shortstack{(1a) \\ create()}};

    \path (observer.south |- robserver_front) coordinate (startCreateRObserver);
    \path (robserver_front.east) coordinate (endCreateRObserver);
    \draw[arrow] (startCreateRObserver) -- (endCreateRObserver)
      node[textnode, midway, above] {\shortstack{(1b) \\ create()}};

    \path (robserver.south) ++(0.0, -0.4) coordinate (startSendRObserver);
    \path (startSendRObserver -| observer.south) coordinate (endSendRObserver);
    \draw[arrow,dashed] (startSendRObserver) -- (endSendRObserver)
      node[textnode, midway, above] {};

    \path (endSendRObserver) ++(0.0, -0.1) coordinate (startSendObserver);
    \path (startSendObserver -| main.south) coordinate (endSendObserver);
    \draw[arrow,dashed] (startSendObserver) -- (endSendObserver)
      node[textnode, midway, above] {};

    \path (main.south |- subprocess.west) coordinate (startCreateSubprocess);
    \path (subprocess.west) coordinate (endCreateSubprocess);
    \draw[arrow] (startCreateSubprocess) -- (subprocess.west)
    node[textnode, midway,
    above] {\shortstack{(2a) \\ create()}};

    \path (startCreateSubprocess) ++(0.0, -1.4) coordinate
    (startSendSendObservers);
    \path (startSendSendObservers -| subprocess.south) coordinate
    (endSendObservers);
    \draw[arrow] (startSendSendObservers) -- (endSendObservers) node[textnode,
    midway, above] {\shortstack{(2b) \\ send(remote\_obs)}};

    \path (startSendSendObservers) ++(0,-1.0) coordinate (startTestToSub);
    \path (startTestToSub -| subprocess.south) coordinate (endTestToSub);
    \draw[arrow] (startTestToSub) -- (endTestToSub) node[textnode,
    midway, above] {\shortstack{(2c) \\ send(test)}};

    \path (thread.west -| subprocess.south) coordinate (startThread);
    \path (thread.west) coordinate (endThread);
    \draw[arrow] (startThread) -- (thread.west) node[textnode, midway,
    above] {\shortstack{(2d) \\ start(test)}};

    \path (startThread) ++(0, -0.5) coordinate (startTestToThread);
    \path (startTestToThread -| thread.south) coordinate (endTestToThread);

    \path (endTestToThread) ++(0,-0.6) coordinate (executeTest);
    \node[textnode] at (executeTest) {\shortstack{(3) \\ execute(test)}};

    \path (executeTest) ++(0.0, -0.5) coordinate (startCallRObservers);
    \path (startCallRObservers -| robserver) coordinate (endCallRObservers);
    \draw[arrow] (startCallRObservers) -- (endCallRObservers)
      node[textnode, midway, above] {\shortstack{(4) \\ notify(test)}};

    \path (endCallRObservers) ++(0.0, -0.2) coordinate (startRCallRObservers);
    \path (startRCallRObservers -| executeTest) coordinate (endRCallRObservers);
    \draw[arrow,dashed] (startRCallRObservers) -- (endRCallRObservers)
      node[textnode, midway, above] {};

    \path (endRCallRObservers) ++(0,-0.8) coordinate (startResultsToSubprocess);
    \path (startResultsToSubprocess -| subprocess.south) coordinate
    (endResults);
    \draw[arrow,dashed] (startResultsToSubprocess) -- (endResults)
    node[textnode, midway,
    above] {\shortstack{(5a) return \\ results}};

    \path (endResults) ++(0.0, -0.2) coordinate (startResultsToMain);
    \path (startResultsToMain -| main.south) coordinate (endResultsToMain);
    \draw[arrow,dashed] (startResultsToMain) -- (endResultsToMain)
    node[textnode, midway, above] {\shortstack{(5b) return \\ results}};

    \path (endResultsToMain) ++(0.0, -0.1) coordinate (startCallObservers);
    \path (startCallObservers -| observer) coordinate (endCallObservers);
    \draw[arrow] (startCallObservers) -- (endCallObservers)
      node[textnode, midway, above] {\shortstack{(6) \\ notify(results)}};

    \path (main.south) ++(0,-9.5) coordinate (lifelineEndMain);
    \begin{scope}[on background layer]
      \draw[dashed] (main.south) -- (lifelineEndMain);
      \draw[dashed] (subprocess.south) -- (startResultsToMain);
      \draw[dashed] (thread.south) -- (startResultsToSubprocess);
      coordinate (lifelineEndObserver);
    \end{scope}

    \draw[dashed] (observer_back.south) -- (lifelineEndMain -| observer_back.south);
    \draw[dashed] (observer.south) -- (lifelineEndMain -| observer.south);
    \draw[dashed] (observer_front.south) -- (lifelineEndMain -| observer_front.south);

    \draw[dashed] (robserver_back.south) -- (lifelineEndMain -| robserver_back.south);
    \draw[dashed] (robserver.south) -- (lifelineEndMain -| robserver.south);
    \draw[dashed] (robserver_front.south) -- (lifelineEndMain -| robserver_front.south);
    
    \begin{scope}[on background layer]
      \path let \p1 = (main.east), \p2 = (subprocess.west) in
        coordinate (sep) at ({0.5*(\x1+\x2)}, 0);
      \draw[dotted, ultra thick] (sep) ++(0,0.3) -- ++(0,-10);
    \end{scope}

  \end{tikzpicture}}
  \caption{\label{fig:subprocess-execution}%
    \Pynguin's \subprocessexec model%
  }
\end{subcaptionblock}
\caption{\label{fig:pynguin-execution-modes}%
  The two execution models' execution sequences%
}
\end{figure}

To address critical issues encountered during test-case
generation, such as \segfaults, we
propose an architectural shift in \pynguin's test-execution model: the
possibility to execute test cases in isolated subprocesses.
This approach effectively mitigates FFI-related errors
by leveraging operating system-level process management (instead of
application-level thread management).
However, this new architecture introduces some challenges and
overheads. Firstly, this approach is slower, as
spawning a subprocess is
more costly than spawning a thread because it requires to start a new
Python interpreter instead of allocating a new thread in the same
process, which is a much lighter operation.
This is aggravated by the
impossibility to use shared memory between processes, requiring an
additional serialisation step to send data between them.
Secondly, \pynguin's architecture is heavily based on
the observer pattern, which causes challenges because tests running in
subprocesses need to be observed from the main process.
We investigate the performance overhead in \cref{sec:rq4_results}.

\subsubsection{Architecture Refactoring}
To add \subprocessexec to \pynguin's execution model we decoupled
the observers from the test-execution threads,
as the original \threadedexec system had a single class that conflated
responsibilities by containing methods that were executed in both the
test-execution thread and the main thread.
While this design was viable in a threaded environment due to
shared memory, it relied on direct state access, which rendered
observer instances non-serialisable. Serialisability, however, is a
prerequisite for transferring objects between processes and is thus
essential for \subprocessexec. Consequently, the original design was
incompatible with process-based isolation, which requires explicit
communication across separate memory spaces.

The new model divides observers into two categories.
\emph{Main Process Observers} run in the main process and monitor high-level
information that is not directly related to test execution, such as the
current iteration of the genetic algorithm or the best test cases found so far.
\emph{Remote Observers} must be sent to the test-executor subprocess
and run there to
collect test execution results, such as coverage information or
assertion results.
With \subprocessexec, as illustrated in \Cref{fig:subprocess-execution},
the \emph{Main Process} creates all observers---both remote~(1b)
and main process
observers~(1a)---and creates a \emph{Test Executor Subprocess}~(2a).
The \emph{Main Process} sends the remote observers~(2b) and the test case~(2c)
to be executed to the \emph{Test Executor Subprocess} which then
starts a \emph{Test Executor Thread}~(2d) to execute the test
case~(3) similar to \threadedexec. We still require the thread for
fine-grained control over the test execution, e.g., to enforce
timeouts. After creating the \emph{results} using the remote observer
data~(4), they are sent back to the main process~(5a, 5b) and used by
the main process observers~(6).
Transferring remote observers and test cases between processes requires
serialisation. For this purpose, we use the
\toolname{dill}~\cite{dill_github}
and \toolname{multiprocess} modules from
\toolname{Pathos}~\cite{McKerns2010, McKerns2011,pathos_github}.
In short, we added the subprocess as an additional intermediate layer
between the main process and the test-execution thread.
In case of a critical C error only the respective subprocess might
crash, leaving the main process, which runs \pynguin, intact.

\subsubsection{Automatic Execution-Mode
Selection}\label{sec:approach:automatic-selection}
Using \subprocessexec is more robust when testing \SUTs that rely on
\FFI code, but it also incurs a runtime overhead compared
to \threadedexec. As \cref{sec:eval} shows, \subprocessexec is
preferable when testing \SUTs that use \FFI code, while
\threadedexec is faster for pure-Python \SUTs.
When running \Pynguin, the user may not know whether a particular \SUT module
uses the FFI and thus might require \subprocessexec or not.
We added three approaches to automatically decide
which execution mode to use.

\textit{\Heuristic}.
The \heuristic execution mode
automatically selects the optimal execution strategy by detecting
FFI code during an initial analysis of the SUT\@.
The detection mechanism first inspects the file extension of
each imported module, identifying \inlinelst{.so} and
\inlinelst{.pyd} files as C-extensions. For \inlinelst{.py} files, it
checks for a pure Python implementation using
\inlinelst{inspect.getsource}~\cite{python_inspect_getsource},
which fails on non-pure-Python modules. If \pynguin detects
C-extensions, it uses \subprocessexec; otherwise, it defaults to
\threadedexec.
A whitelist permits specific built-in modules,
such as \inlinelst{sys} and \inlinelst{math}, to use \threadedexec
despite their underlying C implementations.

\textit{\Fallback}.
The \fallback execution mode uses the faster \threadedexec and restarts
the search with \subprocessexec when there is a crash.
To achieve this, we added a master-worker architecture to
\pynguin. When activated, \Pynguin acts as a master
process that in turn creates a worker in a \subprocess to do the test
generation.
If the worker crashes in \threadedexec due to a critical
C-error, the master adjusts the search time by subtracting used time,
enables \subprocessexec and restarts the search with a new worker.
While this introduces another \subprocess layer, there are only
two communication steps between the master and worker---one at the
worker start and one when the worker finishes. Thus, this does
not result in a notable runtime overhead.

\textit{\Fallbackheuristic}.
When combining \heuristic and \fallback,
\Pynguin does not use \threadedexec as initial execution mode but decides
on it using the heuristic. Upon a crash, the search is restarted with
\subprocessexec.

\subsection{Revealing Faults}
\label{sec:approach:crash-revealing-generation}

By preventing \Pynguin from crashing, \subprocessexec enables the
generation of test cases for modules that use C/\CPP code.
We extended \pynguin to export the test cases that cause a
crash without \subprocessexec separately from the final test suite.
Those \crashrev test cases do not contribute to coverage, but help in
reproducing and debugging the crash. We later execute each exported
\crashrev test case to record the error code which caused the crash.

\CrashRev tests may not be reproducible due to non-deterministic
behaviour or internal state of the \SUT.
When reporting a crash, it is important to provide a
reproducible example to help developers isolate and fix the root cause.
To address this, we introduced a re-execution
step into \pynguin. This step checks whether the generated \crashrev
test case can still trigger the crash.

\section{Evaluation}\label{sec:eval}

We introduced \subprocessexec as described in the previous section.
To evaluate its effectiveness, we ask:
\begin{rqlist}[resume]
\item \label{rq:crashes} \emph{How many \pynguin crashes can be avoided by
  using \subprocessexec?}
\end{rqlist}
\smallskip

By preventing \pynguin from crashing, \subprocessexec allows
generating \crashrev tests which detect faults in the SUT\@.
Since multiple \crashrev tests can be generated for the same fault,
we group the tests based on the last Python statement
called before the crash occurred and the exit code of the
\crashrev test.
We define the last statement as \emph{crash cause}.
To evaluate how well \subprocessexec performs in detecting real-world
faults, we ask:
\begin{rqlist}[resume]
\item \label{rq:unique-crash-causes} \emph{How many unique crash causes can
  be identified using the \crashrev tests generated by \subprocessexec?}
\end{rqlist}
\smallskip

While \subprocessexec allows for testing modules that would otherwise
cause \pynguin to crash, resulting in \qty{0}{\percent} coverage,
it also introduces a performance overhead due to the subprocess
creation and inter-process communication. To examine the overall
impact of this on code coverage, we ask:
\begin{rqlist}[resume]
\item \label{rq:coverage}\emph{What is the impact of using \subprocessexec on
  \linebreak achieved branch coverage?}
\end{rqlist}

\subsection{Evaluation Setup}\label{sec:eval:setup}

We conducted an empirical evaluation as follows to answer our research
questions.

\subsubsection{Subjects}\label{sec:eval:setup:subjects}
Datasets previously used to evaluate \pynguin
did not focus on Python modules that use
C-extensions~\cite{lukasczyk_automated_2020,lukasczyk_pynguin_2022,lukasczyk_empirical_2023,lemieux_codamosa_2023}.
Thus, we created a new dataset, named \emph{\dsc}.

\begin{table}[t]
  \centering
  \caption{Projects, modules per project (M.), version tag,
    \SUT modules size (min, mean, max) in lines of code (LoC),
    and project size in megabytes (MB) for Python, C/\CPP code
  and the percentage of C/\CPP code in the project.}
  \resizebox{\linewidth}{!}{
    \begin{tabular}{l r l rS[table-format=3.0]r SSS}
      \toprule
      \textbf{Project} & \textbf{M.} & \textbf{Tag} &
      \multicolumn{3}{c}{\textbf{\SUT Size (LoC)}} &
      \multicolumn{3}{c}{\textbf{Project Size (MB)}} \\
      \cmidrule(lr){4-6} \cmidrule(lr){7-9}
      & & & \textit{min} & \textit{mean} & \textit{max} &
      \textit{Python} & \textit{C/\CPP} & \textit{\%} \\
      \midrule
      cffi        & 13    & 1.17.1 & 65  & 467.77 & 1294 & 1.0903  &
      0.5840 & 34.76 \\
      coverage    & 38   & 7.8.0  & 35  & 233.63 & 787  & 1.5049  &
      0.0475 & 2.31  \\
      debugpy     & 18   & 1.8.14 & 40  & 210.83 & 782  & 6.2306  &
      0.1012 & 1.50  \\
      distlib     & 11   & 0.3.9  & 107 & 555.45 & 1373 & 1.3403  &
      0.0342 & 2.48  \\
      msgpack     & 1    & 1.1.0  & 79  & 376.00 & 673  & 0.0818  &
      0.0508 & 30.62 \\
      mypy        & 4  & 1.15.0 & 20  & 303.75 & 803  & 5.7418  &
      0.2824 & 4.68  \\
      numba       & 236  & 0.61.2 & 9   & 404.40 & 4791 & 10.2199 &
      0.9892 & 8.81  \\
      numpy       & 83  & 2.2.5  & 7   & 248.78 & 2947 & 10.8358 &
      6.6188 & 37.20 \\
      pandas      & 167  & 2.2.3  & 10  & 548.07 & 5199 & 19.9652 &
      0.3382 & 1.53  \\
      pycparser   & 10   & 2.22   & 42  & 588.70 & 2023 & 0.5104  &
      0.0730 & 12.50 \\
      pygments    & 115  & 2.19.1 & 19  & 370.29 & 3134 & 4.4781  &
      0.1855 & 2.52  \\
      pymongo     & 50   & 4.12.1 & 14  & 352.28 & 1581 & 5.9012  &
      0.1774 & 2.90  \\
      pytz        & 4    & 2025.2 & 86  & 128.75 & 208  & 0.1038  &
      0.2573 & 35.15 \\
      rapidfuzz   & 6   & 3.13.0 & 66  & 145.69 & 374  & 0.4291  &
      0.4000 & 38.38 \\
      scipy       & 8  & 1.15.2 & 14  & 359.00 & 1006 & 19.3281 &
      6.8487 & 22.45 \\
      shapely     & 24   & 2.1.0  & 8   & 100.31 & 447  & 1.0781  &
      0.2602 & 18.94 \\
      simplejson  & 3    & 3.20.1 & 31  & 217.00 & 511  & 0.1674  &
      0.1041 & 38.35 \\
      tensorflow  & 745 & 2.19.0 & 6   & 232.42 & 2683 & 43.6994 &
      97.2581& 56.61 \\
      torch       & 92  & 2.7.0  & 8   & 227.46 & 2310 & 65.9998 &
      41.1158& 35.85 \\
      watchdog    & 15   & 6.0.0  & 20  & 139.47 & 586  & 0.3373  &
      0.0314 & 8.50  \\
      wrapt       & 5    & 1.17.2 & 43  & 183.60 & 443  & 0.2872  &
      0.0986 & 25.53 \\
      \bottomrule
    \end{tabular}
  }
  \label{tab:dataset_stats}
\end{table}

\paragraph*{DS-C}

\Cref{tab:dataset_stats} shows an overview of \dsc,
which contains \modules modules from \libraries popular Python
libraries that use C-extensions.
We used \textsc{cloc}~\cite{cloc_tool}
to measure the size of the used modules in \emph{lines of code}~(LoC)
and the \textsc{GitHub API}~\cite{github_api_languages}
to determine the size of the projects in megabytes for Python,
C/\CPP, and the percentage of C/\CPP code.

The dataset was created as follows:
\begin{inparaenum}[(i)]
\item we collected the \pypiLibsInitial most popular Python libraries from
  the \emph{Python Package Index}~(\pypi) according to the
  \textsc{Top PyPI Packages}~\cite{taskaya_reiz_2021,top_pypi_packages}.
\item For each project, we fetched metadata from \pypi and used the
  \textsc{GitHub API} to identify its GitHub repository and the latest tags.
  Since Git tags often correspond to release versions on \toolname{PyPI},
  we matched the version tag in the \pypi metadata with those from
  GitHub to identify the exact commit used for the release.
\item As we require access to the source code to analyse
  the faults revealed by
  \pynguin, we filtered out projects that are not hosted on GitHub
  or where the tag could not be matched with a commit.
\item In some cases, multiple \toolname{PyPI} packages reference the same
  GitHub repository (e.g., when a single project distributes multiple
  related components such as \emph{tool-core} and \emph{tool-extension}).
  We removed such duplicates based on identical GitHub URLs.
\item To focus on projects that use C-extensions, we queried the
  \textsc{GitHub API} for each project's language composition and retained
  only those with at least \qty{1}{\percent} of C or \CPP code.
\item We searched for all public (\ie, modules that do not start with an
  underscore) and non-test (\ie, modules that do not contain \texttt{test}
  in the name or package name) modules and added them to the dataset.
\item We removed all modules that do not contain any code for \pynguin to test
  (\ie, modules that have no public classes or functions), modules that
  cannot be imported by \pynguin (\eg, due to missing dependencies or
  import errors) and all modules that require coroutines, which \pynguin
  does not support. In total, we removed \removedModules modules
  from the dataset due to these constraints.
\item Some modules are trivial to test. For instance, if a module
  contains only a single function without branching, \pynguin often
  achieves \qty{100}{\percent} branch coverage during the initial random
  generation phase, regardless of the execution mode (\threadedexec or
  \subprocessexec).
  Those trivial modules are neither interesting for measuring
  the performance of \subprocessexec nor for generating \crashrev tests.
  Aligning with previous work~\cite{lemieux_codamosa_2023}, we
  removed \CTrivialModules trivial modules where \pynguin achieved
  \qty{100}{\percent} branch coverage within
  \qty{1}{\minute}. %
\end{inparaenum}

Our final dataset consists of \modules modules from \libraries popular Python
libraries that use C-extensions. Previous work on \pynguin
used datasets with \qtyrange{104}{486}{}
modules~\cite{lukasczyk_automated_2020, lukasczyk_pynguin_2022,
lukasczyk_empirical_2023, lemieux_codamosa_2023}.
We believe our dataset has a good balance
between representativeness (popular projects from different application
domains) and overall execution time of the evaluation. The use of
open-source projects also supports reproducibility and
enables us to further confirm found crashes.

\paragraph*{DS-CodaMosa}
The new \dsc dataset targets evaluating modules using C-extensions.
To evaluate our approach on an unbiased dataset,
we additionally executed \pynguin with and without \subprocessexec on
\dscodamosa, previously used to evaluate
\codamosa~\cite{lemieux_codamosa_2023}.
We decided to use this dataset as it is the largest one
previously used for evaluating \pynguin with \exnum{\codamosaModules}
modules from \num{27} projects.

\subsubsection{Experiment Settings}\label{sec:eval:setup:settings}

We implemented \subprocessexec in \pynguin and executed it on each subject.
We used a \toolname{Docker} container to
isolate the executions from their environment based on Python~3.10.16. All tool
runs were executed on dedicated servers, each equipped with an AMD EPYC 7443P
CPU and \qty{256}{\giga\byte} RAM. However, we only assigned a single
CPU core and \qty{4}{\giga\byte} of RAM to each run to simulate a
more constrained environment, closer to the conditions in which the
SUTs will have to operate, and to be able to save time by executing
multiple runs in parallel.

We used \pynguin's implementation of DynaMOSA~\cite{PKT18b} with the
same parameter settings and a search budget of \qty{600}{\second},
which was effective for test generation~\cite{lukasczyk_empirical_2023}.
To minimise the influence of randomness we
executed \pynguin on each subject \num{30} times with \threaded,
\subprocess, \heuristic, \fallback and \fallbackheuristic mode
each~\cite{AB14}.
Assuming that all runs require the full budget of \qty{600}{\second},
we need
\(5\,\text{configurations} \times \modules\,\text{modules} \times
  \qty{600}{\second} \times 30\,\text{repetitions}=
\qty{\CExecutionTime}{\hour} \)
to execute \pynguin with \dsc and
\(5\,\text{configurations} \times \codamosaModules\,\text{modules} \times
  \qty{600}{\second} \times 30\,\text{repetitions}=
\qty{\CodamosaExecutionTime}{\hour} \)
to execute it with \dscodamosa (not parallelised).

\subsubsection{Experiment Procedure}\label{sec:eval:setup:procedure}

As a basis for answering \ref{rq:crashes} and
\ref{rq:unique-crash-causes} we use
the results of \subprocess
and \threaded runs.
We additionally use the results of
\heuristic, \fallback and \fallbackheuristic runs
when analyzing performance in \ref{rq:coverage}.
For every successful run, we collect the achieved branch coverage
while all crashed runs achieve \qty{0}{\percent} coverage,
as crashed runs do not export any test cases.
When using \subprocessexec, \pynguin exports a \crashrev test
case whenever it encounters a C-related crash during the execution of
the module under test.
With \fallback and \fallbackheuristic mode, and when the first execution
crashes, only the second execution attempt using \subprocessexec
is considered for final coverage, as the first execution did not
result in any test cases. The coverage during the first execution is
set to \qty{0}{\percent}.
We use both, the collected coverage data and the generated
\crashrev tests to answer all research questions~(\ref{rq:crashes},
\ref{rq:unique-crash-causes}, \ref{rq:coverage}).

To address \ref{rq:unique-crash-causes}, we need to ensure that all
crash causes are reproducible. \Pynguin's test
execution, which uses the \toolname{multiprocess}
module for performance reasons,
can lead to \crashrev tests not being generally reproducible.
To resolve this, we re-executed all \crashrev tests with \pytest
using new Python interpreters with the same version
and dependencies.
After re-execution, we removed non-reproducible tests, filtered the
\crashrev tests to only retain those
causing timeouts or crashes when run in new
interpreters, and recorded their exit codes.
We chose to retain \crashrev tests that consistently timed out after
\qty{30}{\second}, as this duration is, in our judgment, sufficient
to complete a typical test generated by \pynguin, and may indicate
faults in the \SUT.
To obtain more precise information about the causes of the crashes
in \crashrev tests,
we analysed the logs produced by Python's \ffiname{faulthandler} module.
These logs provide the traceback of the last Python function calls
prior to the crashes, enabling us to heuristically group tests by the
final Python function executed. We used a heuristic because, in
Python 3.10, the \ffiname{faulthandler} module only provides
traceback dumps at the Python layer, but not at the C layer.
Therefore, we know which Python functions were called
immediately before the crashes, but we cannot determine whether they
have a common or different cause at the C layer. Starting with Python
3.14, the \ffiname{faulthandler} module also provides the C stack
trace, but not all libraries in our dataset were compatible with
this version, so we were unable to use it
for our evaluation. After a manual analysis, we also decided to
exclude some tests that contained calls to 3 types of functions that
put the interpreter in an unstable state, such as the
\ffiname{\_\_del\_\_} functions. These were causing crashes when
executing code that did not have any faults, thereby causing a bias.
For further details, we refer to the replication package~\cite{dataset}.

\subsubsection{Evaluation Metrics}\label{sec:eval:setup:metrics}

For each execution of a module, we keep track of the branch coverage over time
as well as the overall branch coverage at the end of the search.
To evaluate how many crashes \subprocessexec can avoid~(\ref{rq:crashes}),
we count and compare the number of successful executions of \pynguin with
\threadedexec and \subprocessexec.
We measure \pynguin's ability to generate \crashrev
tests~(\ref{rq:unique-crash-causes})
by counting the number of generated \crashrev tests and checking how
many of them are reproducible.
To assess if one configuration performs better
than another configuration in terms of avoided crashes~(\ref{rq:crashes})
and overall coverage~(\ref{rq:coverage}), we
use the Mann-Whitney U-test~\cite{mann_test_1947} at $\alpha =
\exnum{0.05}$. We compute the Vargha and Delaney effect size
\effectsize~\cite{VD00} to investigate the difference in the
achieved overall coverage between two configurations.
These metrics do not make assumptions on the data distribution and are
in line with recommendations for assessing randomised
algorithms~\cite{AB14}
and were used in previous work~\cite{lukasczyk_automated_2020,
  lukasczyk_pynguin_2022, lukasczyk_empirical_2023, lemieux_codamosa_2023,
yang_llm-enhanced_2025}.

\subsection{Threats to Validity}\label{sec:eval-threats}

\textit{Internal Validity}.
While the \subprocessexec~(see \cref{sec:approach:subprocess})
isolates \SUT crashes, the performance overhead of
inter-process communication affects achieved
coverage. Changes in performance may stem from the execution overhead
rather than from crash isolation.
Nevertheless, \subprocessexec
enables testing of previously untestable modules, improving over
\threadedexec. Furthermore,
we added a \heuristic %
which combines the benefits of both approaches.

Another threat is that some crashes found by generated \crashrev test
cases are not reproducible upon re-execution.
Non-reproducible crashes may be falsely counted as
faults, leading to an overestimation of fault-detection effectiveness.
This behaviour is often due to non-determinism or the
internal state of the \SUT. Other factors might be
the internal state of the execution environment or side effects introduced
by \pynguin's instrumentation mechanisms. Such, so-called \emph{flaky tests},
are a known issue for test generation tools~\cite{gruber_automatic_2024} and
do not specifically occur because of \subprocessexec.
\smallskip

\textit{External Validity}.
The generalisability of our findings may be limited. \dsc consists of
\modules modules from \libraries projects on \pypi, selected to
include the most popular libraries with C-extensions and filtered
using the GitHub API and specific criteria (\eg, the presence of
C/\CPP code).
These design choices may introduce selection bias, potentially
exclude projects or modules where \subprocessexec behaves
differently, and limit the transferability of the results to other
Python libraries or \CExtension modules.
The bugs identified are also tied to these libraries,
and the prevalence of such bugs
detectable by this method might vary.
We mitigate this threat by evaluating \pynguin also on the
\dscodamosa dataset, which is unbiased and does not rely on the
presence of C/\CPP code.

Another possible threat regarding our \dsc dataset is that we only
selected modules that are public according to Python, meaning their names do
not start with an underscore. This approach removes
modules that developers do not want to be used directly, and those
that remain do not necessarily correspond to the public API defined
in the documentation of the projects. There exists, however, no reliable way to
automatically differentiate between public and private APIs in Python.
We limited the dataset to public modules
of projects,
because the Python documentation for C-extension
modules~\cite{python_extension_modules,
python_errors_exceptions}
states that public functions
defined in C must raise an exception in the case of a failure
and not crash the interpreter.
Thus, we can be sure that if a crash occurs,
unless the implementation violates the Python documentation, there
is either a bug in a public C-function or in the
Python code calling a private C-function.

We do not compare \pynguin with other test
generation tools, which introduces another threat to validity.
However, to the best of our knowledge, there is no
existing tool that can handle C-exceptions. Therefore, a comparison
to other tools would not help to understand the improvements achieved by
\subprocessexec.
We instead focus on the comparison of \pynguin with and without
\subprocessexec. %
\smallskip

\textit{Construct Validity}.
The metrics used to evaluate the improvements focus on the crash
avoidance of \pynguin~(\ref{rq:crashes}),
the detection of crash causes~(\ref{rq:unique-crash-causes})
and the impact on branch coverage~(\ref{rq:coverage}).
When measuring branch coverage, we consider crashed runs to achieve
\qty{0}{\percent} coverage. This couples robustness gains with search
effectiveness, which is intentional: crashed runs do not generate tests and
therefore do not contribute to coverage.
While these aspects are important indicators of bug-finding
capability and robustness, they may not comprehensively capture all
features of test case quality, such as the ability to detect
non-crashing logical errors, the diversity of inputs, or the
maintainability of the generated tests.
However, we focus on \crashrev tests, as they are
the most relevant for our approach and leave the other aspects for
future work.
Furthermore, the evaluation does not deeply analyse the severity
or types of these crashes, which could provide a more nuanced
understanding of the practical impact. Analyzing the severity
is an important, yet a different problem, and thus we also
leave it for future work.
Finally, unique crash causes are identified using a heuristic approximation
based on the last Python function called before the crash. Although
this method is not perfect and may not accurately reflect the true
native-level root causes of the crashes, it provides a first simple
automated estimation. We acknowledge that this is a limitation of our
approach and that a more accurate method using the
\ffiname{faulthandler} module of Python 3.14 could be used in future
work to better classify the crash causes, once all SUTs support Python~3.14.

\begin{figure}[!t]
  \centering
  \includegraphics{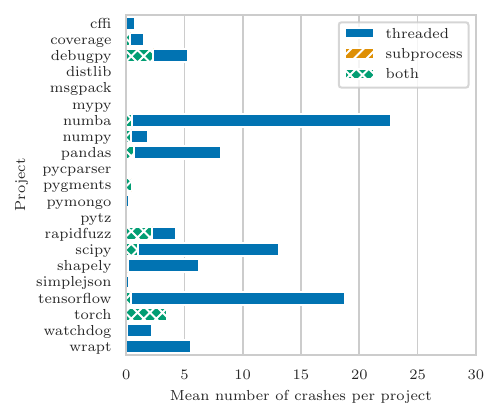} %
  \caption{Distribution of crashes when using
  \threadedexec and \subprocessexec on \dsc.}
  \label{fig:c-crash-distribution}
\end{figure}

\subsection{RQ1: Avoiding Crashes}\label{sec:rq1_results}

\begin{table}[!t]
  \centering
  \caption{Number of modules where \subprocessexec (treatment)
    had (significantly) fewer, equal, or (significantly) more crashes than
    \threadedexec (control) and the Vargha and Delaney
  \effectsize~effect size for \dsc.}
  \begin{tabular}{@{}lrrrS@{}}
  \toprule
  \textbf{Project} & \textbf{Fewer (Sig)} & \textbf{Equal} &
  \textbf{More (Sig)} & \textbf{\effectsize} \\
  \midrule
  cffi & 2 (1) & 11 & 0 (0) & 0.512821 \\
  coverage & 6 (2) & 30 & 2 (1) & 0.519737 \\
  debugpy & 7 (4) & 6 & 5 (1) & 0.549074 \\
  distlib & 2 (0) & 9 & 0 (0) & 0.503030 \\
  msgpack & 0 (0) & 1 & 0 (0) & 0.500000 \\
  mypy & 0 (0) & 4 & 0 (0) & 0.500000 \\
  numba & 200 (184) & 29 & 7 (0) & 0.874364 \\
  numpy & 21 (7) & 59 & 3 (0) & 0.530120 \\
  pandas & 129 (87) & 36 & 2 (0) & 0.622655 \\
  pycparser & 0 (0) & 10 & 0 (0) & 0.500000 \\
  pygments & 5 (3) & 105 & 5 (2) & 0.503913 \\
  pymongo & 7 (2) & 43 & 0 (0) & 0.505000 \\
  pytz & 0 (0) & 4 & 0 (0) & 0.500000 \\
  rapidfuzz & 1 (1) & 4 & 1 (1) & 0.497222 \\
  scipy & 8 (5) & 0 & 0 (0) & 0.729167 \\
  shapely & 22 (13) & 2 & 0 (0) & 0.614583 \\
  simplejson & 1 (0) & 2 & 0 (0) & 0.505556 \\
  tensorflow & 624 (609) & 113 & 8 (0) & 0.820895 \\
  torch & 24 (14) & 31 & 37 (26) & 0.497645 \\
  watchdog & 2 (2) & 12 & 1 (0) & 0.535556 \\
  wrapt & 1 (1) & 4 & 0 (0) & 0.593333 \\
  \midrule
  \textbf{Total} & \textbf{1062 (935)} & \textbf{515} & \textbf{71
  (31)} & \textbf{0.717} \\
  \bottomrule
\end{tabular}

  \label{tab:crash-sig-summary}
\end{table}

Recall that, in the context of this paper, a crash refers to a fault
that causes the Python interpreter to terminate unexpectedly,
interrupting test generation, preventing coverage
progress, and the discovery of additional faults.
To determine how many crashes \pynguin can avoid by using
\subprocessexec, we investigate the number of crashes that
\pynguin encounters when using \threadedexec and \subprocessexec.
\Cref{fig:c-crash-distribution} shows the number of crashes averaged
over all modules per project for \pynguin on \dsc.
\Cref{tab:crash-sig-summary} additionally shows the number of modules
where \subprocessexec had (significantly) less, equal, or (significantly)
more crashes than \threadedexec and the Vargha and Delaney
\effectsize~effect size for an in-depth comparison.

\textit{Fewer Crashes with \SubprocessExec}.
On projects that rely heavily on C/\CPP extensions for
performance-critical operations, such as the machine learning libraries
\toolname{numba} and \toolname{tensorflow}, \threadedexec encounters
many crashes while \subprocessexec avoids most of them.
\Subprocessexec produced (significantly) fewer
crashes than \threadedexec in \exnum{200} modules (sig: \exnum{184})
and \exnum{624} modules (sig: \exnum{609}) respectively.
The mean \effectsize~effect sizes of \num{0.874} and
\num{0.821} indicate a large positive effect.
We conjecture this is because \subprocessexec prevents the interpreter
from crashing by isolating the \SUT in a separate process.

\textit{No/Few Execution Model Crashes}.
Some projects (\eg the packaging library \toolname{distlib}
or the serialization library \toolname{msgpack})
show no crashes under either execution model.
Others (\eg the syntax highlighter \toolname{pygments} or
the database library \toolname{pymongo}) exhibit very few crashes.
These projects may be well-tested, exercise only pure-Python APIs,
or \pynguin may not trigger the native-code paths
that tend to cause interpreter-level failures.
These projects have in common that they are primarily written in pure
Python and use less C code.

\textit{Both Execution Models Crash}.
Some crashes occur regardless of the execution mode.
Those indicate limitations inherent to the
\pynguin tool, such as its inability to handle coroutines,
and that \subprocessexec does not address. For example,
projects like the automated release tool \toolname{python-semantic-release} or
the project scaffolding tool \toolname{cookiecutter} require
specific environmental conditions to run correctly, such as
the presence of a version-controlled repository, specific
configuration files, or network connectivity. \Pynguin is currently
not able to emulate these conditions, leading to
crashes regardless of the isolation mode used.
As these are not related to the execution isolation
provided by \subprocessexec, we pose addressing them as future work.

\textit{More Crashes with \SubprocessExec}.
For a few modules, primarily within the \pyTorch library,
\subprocessexec shows more crashes than \threadedexec.
Manual inspection indicates these are due to infrastructure issues:
we allocated a total of \qty{10}{\hour} for \exnum{30} repetitions
for each module to account \qty{10}{\minute} for dependency installation
and \qty{10}{\minute} for test generation.
Installing \pyTorch includes large dependencies such as
\toolname{CUDA} binaries which took more than \qty{10}{\minute} on average.
This lead to container timeouts depending on the current
infrastructure load and affected both execution models.
These cases do not indicate a flaw in \subprocessexec but are
artefacts of the execution environment.

Overall, the results show an improvement with \subprocessexec, which
increased the number of successful, crash-free modules from \CDefaultSuccess
to \CSubprocessSuccess. This corresponds to a \CCrashesReducedPercent
relative reduction in execution failures.
The superior isolation of \subprocessexec was especially evident with
\CModulesOnlyInSubprocess modules, for which \pynguin achieved at
least one successful run, while it consistently failed to generate test
cases for all \exnum{30} runs when using \threadedexec.
The mean Vargha and Delaney effect sizes~\effectsize of \num{0.717}
indicates a large positive effect of \subprocessexec over \threadedexec.

\begin{summary}{\hypersetup{linkcolor=white}\ref{rq:crashes}: Avoiding Crashes}
  We can avoid up to \CCrashesReducedPercent of all crashes by using
  \subprocessexec. The exact number of avoided crashes depends on the
  project.
\end{summary}

\subsection{RQ2: Unique Crash Causes}\label{sec:rq2_results}

As shown in \ref{rq:crashes}, \subprocessexec can prevent Pynguin's
process from crashing, but it can also store the tests that would
have caused the crashes to help developers reproduce them. We call
those \crashrev tests, and \pynguin generated
\CrashRevealingTotalSubprocess of them. However, some tests no longer
caused crashes when re-executed, which decreased the total of
reproducible tests to \CrashRevealingReproducibleCountSubprocess.
After grouping them using their last Python
instructions called before crashing and their exit codes,
we found \CrashRevealingCauseCount unique crash causes.
Based on this, we identified \identifiedFaults previously
unknown faults and reported them to the respective development
teams. Most of these faults have been confirmed by the developers and
included in the projects' development processes, while others have
not yet been reviewed or have been classified as “won't fix.” Due to
a lack of time and resources from the developers, only a few
have been fixed so far.
Since quantitative data does not provide any insights into the severity
of the detected faults, we classified the crashes based on their exit
codes, as shown in \cref{tab:crash-classification}.
\Pynguin supports a subset of all POSIX signal exit
codes~\cite{linux_man_signal_7} that may be
raised due to programming errors: \segfaults, \aborteds,
\illegalinstructions, \buserrors, \longoomkills, and \fpes.

\begin{table}[!t]
  \centering
  \caption{Distribution of fault types among unique crash causes.}
  \begin{footnotesize}
    \begin{tabular}{@{}lrrrr@{}}
  \toprule
  \textbf{Crash reason} & \textbf{Exit code} & \textbf{Signal} &
  \textbf{Count} & \textbf{Percentage} \\
  \midrule
  Aborted & \exnum{-6} & SIGABRT & \AbortedCount & \AbortedPercent \\
  Segmentation fault & \num{-11} & SIGSEGV & \SegmentationFaultCount
  & \SegmentationFaultPercent \\
  Timeout & None & N/A & \TimeoutCount & \TimeoutPercent \\
  \bottomrule
\end{tabular}

  \end{footnotesize}
  \label{tab:crash-classification}
\end{table}

\begin{figure}[!t]
  \centering
\begin{lstlisting}[
    style=numberedlst,%
    frame=tb,%
  ]
import tensorflow.python.eager.context as module_0

def test_case_0():
    none_type_0 = None
    var_0 = module_0.initialize_logical_devices()
\end{lstlisting}
  \caption{Invoking \inlinelst{initialize_logical_devices} causes a
  \segfault.}
  \label{fig:single-statement}
\end{figure}

\begin{figure}[!t]
  \centering
\begin{lstlisting}[
    style=numberedlst,%
    frame=tb,%
  ]
import tensorflow.python.eager.tape as module_0


def test_case_0():
    list_0 = []
    tape_0 = module_0.Tape(list_0)
    module_0.push_tape(tape_0)
    module_0.variable_accessed(tape_0)
\end{lstlisting}
  \caption{Invoking \inlinelst{variable_accessed} with an invalid
    argument and after setting up the internal state of the \SUT
  causes a \segfault.}
  \label{fig:specific-state}
\end{figure}

\textit{\SegFaults are Predominant}.
\Segfaults occur when the \SUT attempts to access memory outside its
allocated range, causing the OS to terminate the process. Since C
provides no built-in
safeguards against such behaviour, we expected \segfaults to be the
most common crash type and we found \SegmentationFaultCount of them.
We also detected \AbortedCount \aborteds which usually stem from C assertions.
Developers use such assertions to catch bugs as early as possible, by
checking preconditions or invariants. Since they are a faily
widespread practice, this is expected.
These \aborteds should, however, never reach the Python level because
developers are expected to raise Python exceptions instead of letting
the interpreter crash,
as discussed in \cref{sec:eval-threats}.
Finally, we observed \TimeoutCount \timeouts, as defined in
\cref{sec:eval:setup:procedure}. We tried to re-execute them for a
longer time, but they still did not terminate. They may indicate an
infinite loop
or a deadlock in the \SUT and are thus noteworthy.

\textit{Automated State-Dependent Fault Detection}. In certain cases,
such as in \cref{fig:single-statement}, fuzzing might find the same
bugs as our approach by calling the function with random arguments.
In this case, a \segfault is raised from a public module if the
computer does not have a tensor processing unit, TPU\@. Nonetheless,
\pynguin is able to automatically detect such straightforward faults
as well as more complex cases that otherwise require domain knowledge
and manually configuring fuzzers. For example, in
\cref{fig:specific-state}, the internal state of the \SUT is first
initialised and then the \inlinelst{variable_accessed} function is
called with an invalid argument, causing a crash. Although fuzzers
can invoke the function, the crash occurs only in a special \SUT
state. This requires manually configuring the fuzzer.
In contrast, \pynguin reveals these crashes fully automatically
requiring neither configuration nor domain knowledge.

\textit{Non-Reproducible \CrashRev Tests}.
\CrashRevealingNonReproducibleSubprocess of
\CrashRevealingTotalSubprocess detected crashes could not be
reproduced when executing the exported test cases. It is unlikely
that there was no crash in the first place or else \pynguin would not
have exported them. \cref{sec:eval-threats} explains several reasons
for this behaviour. The most noteworthy case is that we
had to remove the few \oomkills because they mainly came from
\toolname{numpy}, which was trying to allocate arrays of
dozens of GB in the generated \crashrev tests, inevitably causing
false positives.

\begin{summary}{\hypersetup{linkcolor=white}\ref{rq:unique-crash-causes}:
  Unique Crash Causes}
  Using \subprocessexec, we detected \CrashRevealingCauseCount unique
  crash causes and identified \identifiedFaults unknown faults. The
  crashes were predominantly \segfaults (\SegmentationFaultPercent),
  with smaller proportions of \aborteds from C assertion violations
  (\AbortedPercent) and \timeouts (\TimeoutPercent).
\end{summary}

\subsection{RQ3: Branch Coverage}\label{sec:rq4_results}

\begin{figure*}[!t]
  \centering
  \begin{subfigure}{0.48\linewidth}
    \centering
    \includegraphics[height=45.1mm]{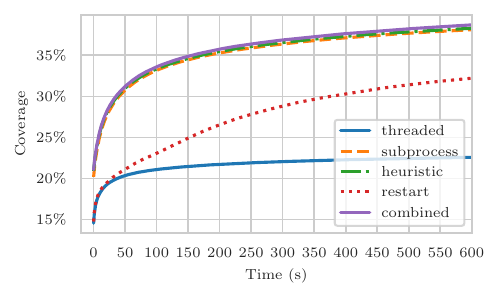}
    \caption{Branch coverage over time on \dsc.}
    \label{fig:c-coverage-over-time-combined}
  \end{subfigure}
  \hfill
  \begin{subfigure}{0.48\linewidth}
    \centering
    \includegraphics[height=45.1mm]{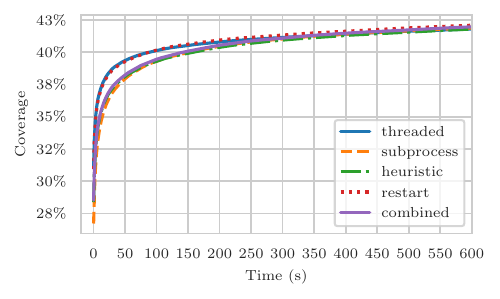}
    \caption{Branch coverage over time on \dscodamosa.}
    \label{fig:codamosa-coverage-over-time-combined}
  \end{subfigure}

  \caption{Branch coverage over time when running \pynguin.
  }
  \label{fig:combined-coverage-over-time}
\end{figure*}
\begin{table*}[t]
  \begin{subtable}{0.48\linewidth}
    \caption{\label{tab:c-coverage-significance}%
      Number of modules and effect size for \dsc.
    }
    \centering
\begin{tabular}{@{}lrrrS@{}}
  \toprule
  \textbf{Treatment} & \textbf{Better (Sig)} & \textbf{Equal} &
  \textbf{Worse (Sig)} & \textbf{\effectsize} \\
  \midrule
  \fallbackheuristic & 1109 (986) & 258 & 281 (134) & 0.716373 \\
  \heuristic & 1111 (976) & 254 & 283 (147) & 0.713324 \\
  \subprocess & 1072 (987) & 245 & 331 (247) & 0.693469 \\
  \fallback & 1075 (675) & 325 & 248 (34) & 0.629184 \\
  \textbf{\threaded} & 0 (0) & 1648 & 0 (0) & 0.500000 \\
  \bottomrule
\end{tabular}

  \end{subtable}
  \hfill
  \begin{subtable}{0.48\linewidth}
    \caption{\label{tab:codamosa-coverage-significance}%
      Number of modules and effect size for \dscodamosa.
    }
    \centering
\begin{tabular}{@{}lrrrS@{}}
  \toprule
  \textbf{Treatment} & \textbf{Better (Sig)} & \textbf{Equal} &
  \textbf{Worse (Sig)} & \textbf{\effectsize} \\
  \midrule
  \fallback & 74 (10) & 349 & 63 (1) & 0.503963 \\
  \textbf{\threaded} & 0 (0) & 486 & 0 (0) & 0.500000 \\
  \fallbackheuristic & 60 (30) & 266 & 160 (112) & 0.444975 \\
  \heuristic & 48 (24) & 270 & 168 (111) & 0.441868 \\
  \subprocess & 53 (37) & 234 & 199 (161) & 0.418208 \\
  \bottomrule
\end{tabular}

  \end{subtable}
  \caption{Number of modules where each treatment configuration~(Treatment)
    performed (significantly) better, equal or (significantly) worse
    than \threadedexec (control) and the Vargha and Delaney
  \effectsize~effect size.}
  \label{fig:coverage-significance}
\end{table*}

\Cref{fig:combined-coverage-over-time} and \cref{fig:coverage-significance}
show the performance differences
between \threaded, \subprocess, \heuristic, \fallback
and \fallbackheuristic mode for \dsc and \dscodamosa.

\textit{Higher Coverage with \SubprocessExec}.
On \dsc, \pynguin with
\subprocessexec achieves a higher branch coverage than with
\threadedexec~(\cref{fig:c-coverage-over-time-combined}).
When using the \heuristic to automatically select the
execution model and when combining this with \fallback~(\fallbackheuristic),
\pynguin achieves an even higher branch coverage.
Solely using \fallback results in coverage similar to \threadedexec
initially and later improving towards \subprocessexec.
Our observations hold for the overall branch coverage,
and these differences are statistically significant,
as \cref{tab:c-coverage-significance} shows.
On \CSubprocessBetter~(\CSubprocessBetterSig significant) modules,
\subprocessexec achieves a (significantly) higher branch coverage than with
\threadedexec, while on \CSubprocessWorse~(\CSubprocessWorseSig significant)
modules \threadedexec achieves a (significantly) higher branch
coverage than with \subprocessexec.
The mean Vargha and Delaney~\effectsize effect size confirms a
medium positive effect of \CSubprocessEffectSize.
Using the \heuristic to select the execution model and combining it
with \fallback~(\fallbackheuristic) further increases the average performance.

Overall, \subprocessexec is better than \threadedexec and \fallbackheuristic
performs best on \dsc.
However, as this increase is heavily influenced by those cases where
\Pynguin with \threadedexec crashes and thus results in
\qty{0}{\percent} coverage, we also investigate how branch coverage
results generalize on \dscodamosa.

\textit{Lower Coverage with \SubprocessExec}.
Even though \cref{fig:codamosa-coverage-over-time-combined} suggests
that \pynguin with \threadedexec achieves similar coverage to
\subprocessexec on \dscodamosa,
\cref{tab:codamosa-coverage-significance} shows that
\pynguin with \threadedexec performs better than with
\subprocessexec on most modules in \dscodamosa.
On \CodamosaSubprocessBetter~(\CodamosaSubprocessBetterSig significant)
modules, \subprocessexec achieves significantly higher branch coverage than
with \threadedexec, while on
\CodamosaSubprocessWorse~(\CodamosaSubprocessWorseSig significant)
modules \threadedexec achieves (significantly) higher branch
coverage than with \subprocessexec.
The mean Vargha and Delaney~\effectsize effect size confirms a small
negative effect of \CodamosaSubprocessEffectSize.
Overall, \threadedexec is better than \subprocessexec
and \fallback mode performs best on \dscodamosa.

\textit{Strategy to Use}.
The best test execution strategy depends on
whether the \SUT uses \FFI modules that might trigger C-related errors.
As this is generally unknown in advance, we implemented
three approaches to automatically decide
which execution mode to use (see \cref{sec:approach:automatic-selection}).
The \heuristic selects \subprocessexec for \CHeuristicSubprocessPercent
of \dsc modules and \CodamosaHeuristicSubprocessPercent of \dscodamosa
modules.
The high rate on \dscodamosa stems from the \heuristic's conservative
design, which chooses \subprocessexec for any module where C-extensions
are imported, regardless of usage or potential to crash.
Restart was triggered
on \CRestartTriggeredModulesPercent of \dsc modules
and on \CodamosaRestartTriggeredModulesPercent of \dscodamosa modules.
\Cref{tab:c-coverage-significance} shows that the \fallbackheuristic
mode performs best on \dsc, achieving a large mean Vargha and
Delaney~\effectsize
effect size of \num{0.716} compared to \threadedexec.
On \dscodamosa, \cref{tab:codamosa-coverage-significance} shows that
only the \fallback mode performs better than \threadedexec, achieving
a negligible mean Vargha and Delaney~\effectsize effect size of \num{0.504}.
Overall, we recommend using \fallback mode, as it performs better than
\threadedexec on both datasets and adds the stability of \subprocessexec
only when needed.

\begin{summary}{\hypersetup{linkcolor=white}\ref{rq:coverage}:
  Branch Coverage}
  \Subprocessexec achieves lower branch coverage than \threadedexec
  in general.
  However, on modules that use C-extensions, \subprocessexec
  outperforms \threadedexec.
  Restarting \pynguin with \subprocessexec only when
  \threadedexec crashes (\fallback) yields the best performance trade-off.
\end{summary}

\section{Related Work}\label{sec:relatedwork}

Automated fault detection typically relies on crash or exception
oracles,
common in fuzzing~\cite{zhu_fuzzing_2022}
and SBSE tools like \evosuite~\cite{fraser_evosuite_2013, fraser_1600_2015}.
However, unlike crash reproduction techniques that require existing
stack traces~\cite{rosler_reconstructing_2013, soltani_guided_2017,
derakhshanfar_botsing_2021, bergel_beacon_2021},
our approach discovers and generates reproducible \crashrev tests
from scratch using only the \SUT.
Furthermore, \pynguin serves as a general-purpose tool,
avoiding the domain-specific constraints,
such as used for RESTful testing~\cite{arcuri_restful_2019}
or the static analysis dependencies of API misuse
detection~\cite{kechagia_effective_2019}.
Finally, while most prior research targets Java or focuses on static
analysis of Python C-extensions~\cite{hu_pythonc_2020},
to the best of our knowledge,
this is the first large-scale study targeting dynamic fault detection
for Python C-extensions.

\section{Conclusions and Future Work}\label{sec:conclusion}

We introduced \subprocessexec to automatically generate \crashrev
tests for Python modules that use C-extensions.
By executing the \SUT in a separate process,
\subprocessexec captures crashes,
generates \crashrev tests
and allows the test generation to continue after a crash
caused by a C-extension module.
Our evaluation on \modules modules
from \libraries libraries shows that
our approach avoids up to \CCrashesReducedPercent of
crashes during test generation.
It produced \CrashRevealingReproducibleCountSubprocess \crashrev
tests, allowing us to identify \identifiedFaults unknown faults.

A large proportion of the bugs found with our approach stem from
missing validation of arguments passed to C functions, which can
cause crashes in C due to unmet assumptions.
To address this, future work should measure code
coverage not only in the Python layer but also in the underlying FFI
code, and enhance \pynguin's fitness function with this information.

Additionally, future work can leverage \subprocessexec to implement
parallel test execution. This would address one of the
bottleneck of test generation---fitness evaluation---and may
overcome the performance overhead of \subprocessexec.

Finally, while our approach focuses on testing the Python projects that
use C-extensions, the underlying idea of executing the \SUT in a separate
process is not limited to Python. It can be applied to any programming
language with process isolation.

\section{Acknowledgment}

This research was partially funded by the CyberExcellence by
DigitalWallonia project (No. 2110186) funded by the Public Service of
Wallonia (SPW Recherche),
and with the support of Wallonie-Bruxelles International (RealM project),
as well as by the German Research Foundation (DFG) under grant FR
2955/5-1 (TYPES4STRINGS: Types For Strings).

\balance
\bibliographystyle{IEEEtran}
\bibliography{main}

\begin{thebibliography}{10}
\providecommand{\url}[1]{#1}
\csname url@samestyle\endcsname
\providecommand{\newblock}{\relax}
\providecommand{\bibinfo}[2]{#2}
\providecommand{\BIBentrySTDinterwordspacing}{\spaceskip=0pt\relax}
\providecommand{\BIBentryALTinterwordstretchfactor}{4}
\providecommand{\BIBentryALTinterwordspacing}{\spaceskip=\fontdimen2\font plus
\BIBentryALTinterwordstretchfactor\fontdimen3\font minus \fontdimen4\font\relax}
\providecommand{\BIBforeignlanguage}[2]{{%
\expandafter\ifx\csname l@#1\endcsname\relax
\typeout{** WARNING: IEEEtran.bst: No hyphenation pattern has been}%
\typeout{** loaded for the language `#1'. Using the pattern for}%
\typeout{** the default language instead.}%
\else
\language=\csname l@#1\endcsname
\fi
#2}}
\providecommand{\BIBdecl}{\relax}
\BIBdecl

\bibitem{stackoverflow_dev_survey_2023}
\BIBentryALTinterwordspacing
{Stack Overflow}. (2023) {Stack Overflow Developer Survey 2023: Most popular technologies}. Accessed: 2025-07-17. [Online]. Available: \url{https://survey.stackoverflow.co/2023/#most-popular-technologies-language-prof}
\BIBentrySTDinterwordspacing

\bibitem{numpy}
\BIBentryALTinterwordspacing
{NumPy Developers}. (2025) {NumPy}. Accessed: 2025-07-01. [Online]. Available: \url{https://numpy.org/}
\BIBentrySTDinterwordspacing

\bibitem{pandas}
\BIBentryALTinterwordspacing
{Pandas Developers}. (2025) {Pandas}. Accessed: 2025-07-01. [Online]. Available: \url{https://pandas.pydata.org/}
\BIBentrySTDinterwordspacing

\bibitem{tensorflow}
\BIBentryALTinterwordspacing
{TensorFlow Developers}. (2025) {TensorFlow}. Accessed: 2025-07-01. [Online]. Available: \url{https://www.tensorflow.org/}
\BIBentrySTDinterwordspacing

\bibitem{scipy}
\BIBentryALTinterwordspacing
{SciPy Developers}. (2025) {SciPy}. Accessed: 2025-07-12. [Online]. Available: \url{https://scipy.org/}
\BIBentrySTDinterwordspacing

\bibitem{cython_tool}
\BIBentryALTinterwordspacing
{Cython Developers}. (2025) {Cython: C-Extensions for Python}. Accessed: 2025-07-01. [Online]. Available: \url{https://cython.org/}
\BIBentrySTDinterwordspacing

\bibitem{cython_compiler_directives}
\BIBentryALTinterwordspacing
------. (2025) {Cython User guide: Compiler directives}. Accessed: 2025-07-01. [Online]. Available: \url{https://cython.readthedocs.io/en/3.1.x/src/userguide/source_files_and_compilation.html#compiler-directives}
\BIBentrySTDinterwordspacing

\bibitem{mitchell_leakbot_2003}
N.~Mitchell and G.~Sevitsky, ``{LeakBot}: An automated and lightweight tool for diagnosing memory leaks in large java applications,'' in \emph{Proc.\ ECOOP}, ser. LNCS, vol. 2743.\hskip 1em plus 0.5em minus 0.4em\relax Springer, 2003, pp. 351--377.

\bibitem{nethercote_valgrind_2007}
N.~Nethercote and J.~Seward, ``Valgrind: A framework for heavyweight dynamic binary instrumentation,'' in \emph{Proc.\ PLDI}.\hskip 1em plus 0.5em minus 0.4em\relax {ACM}, 2007, pp. 89--100.

\bibitem{hu_pythonc_2020}
M.~Hu and Y.~Zhang, ``The {Python}/{C} {API}: Evolution, usage statistics, and bug patterns,'' in \emph{Proc.\ SANER}, 2020, pp. 532--536.

\bibitem{yu_dynamic_2021}
B.~Yu, C.~Tian, N.~Zhang, Z.~Duan, and H.~Du, ``A dynamic approach to detecting, eliminating and fixing memory leaks,'' \emph{J.\ Comb.\ Optim.}, vol.~42, no.~3, pp. 409--426, 2021.

\bibitem{memray_github}
\BIBentryALTinterwordspacing
{Bloomberg}. (2025) {Memray}. Accessed: 2025-06-16. [Online]. Available: \url{https://github.com/bloomberg/memray}
\BIBentrySTDinterwordspacing

\bibitem{Pacheco2007}
C.~Pacheco and M.~D. Ernst, ``Randoop: Feedback-directed random testing for {Java},'' in \emph{Proc.\ OOPSLA Companion}.\hskip 1em plus 0.5em minus 0.4em\relax {ACM}, 2007, pp. 815--816.

\bibitem{fraser_evosuite_2011}
G.~Fraser and A.~Arcuri, ``{EvoSuite}: Automatic test suite generation for object-oriented software,'' in \emph{Proc.\ ESEC/FSE}.\hskip 1em plus 0.5em minus 0.4em\relax {ACM}, 2011, pp. 416--419.

\bibitem{lukasczyk_pynguin_2022}
S.~Lukasczyk and G.~Fraser, ``{Pynguin}: Automated unit test generation for {Python},'' in \emph{Proc.\ ICSE Companion}, 2022, pp. 168--172.

\bibitem{fraser_evosuite_2013}
G.~Fraser and A.~Arcuri, ``{EvoSuite}: On the challenges of test case generation in the real world,'' in \emph{Proc.\ ICST}, 2013, pp. 362--369.

\bibitem{fraser_1600_2015}
------, ``1600 faults in 100 projects: automatically finding faults while achieving high coverage with {EvoSuite},'' \emph{Empir.\ Softw.\ Eng.}, vol.~20, no.~3, pp. 611--639, 2015.

\bibitem{pytest}
\BIBentryALTinterwordspacing
{pytest Developers}. (2025) {pytest Documentation}. Accessed: 2025-06-27. [Online]. Available: \url{https://docs.pytest.org/en/stable/}
\BIBentrySTDinterwordspacing

\bibitem{zhu_fuzzing_2022}
X.~Zhu, S.~Wen, S.~Camtepe, and Y.~Xiang, ``Fuzzing: {A} {Survey} for {Roadmap},'' \emph{{ACM} Comput.\ Surv.}, vol.~54, no.~11, pp. 230:1--230:36, 2022.

\bibitem{lukasczyk_automated_2020}
S.~Lukasczyk, F.~Kroiß, and G.~Fraser, ``Automated unit test generation for {Python},'' in \emph{Proc.\ SSBSE}, ser. LNCS, vol. 12420.\hskip 1em plus 0.5em minus 0.4em\relax Springer, 2020, pp. 9--24.

\bibitem{lukasczyk_empirical_2023}
------, ``An empirical study of automated unit test generation for {Python},'' \emph{Empir.\ Softw.\ Eng.}, vol.~28, no.~2, p.~36, 2023.

\bibitem{lemieux_codamosa_2023}
C.~Lemieux, J.~P. Inala, S.~K. Lahiri, and S.~Sen, ``{CodaMosa}: Escaping coverage plateaus in test generation with pre-trained large language models,'' in \emph{Proc.\ ICSE}, 2023, pp. 919--931.

\bibitem{dataset}
\BIBentryALTinterwordspacing
L.~Berg, L.~Krodinger, S.~Lukasczyk, A.~Panichella, G.~Fraser, W.~Vanhoof, and X.~Devroey, ``{Artefact for the paper "Real-World Fault Detection for C-Extended Python Projects with Automated Unit Test Generation" },'' Mar. 2026. [Online]. Available: \url{https://doi.org/10.5281/zenodo.18879486}
\BIBentrySTDinterwordspacing

\bibitem{jni_docs}
\BIBentryALTinterwordspacing
{Oracle}. (2025) {Java Native Interface (JNI)}. Accessed: 2025-07-01. [Online]. Available: \url{https://docs.oracle.com/javase/8/docs/technotes/guides/jni/}
\BIBentrySTDinterwordspacing

\bibitem{jna_docs}
\BIBentryALTinterwordspacing
{Java Native Access (JNA) Contributors}. (2025) {Java Native Access (JNA)}. Accessed: 2025-07-01. [Online]. Available: \url{https://github.com/java-native-access/jna}
\BIBentrySTDinterwordspacing

\bibitem{haskell_ffi}
\BIBentryALTinterwordspacing
{HaskellWiki contributors}. (2025) {Foreign Function Interface}. Accessed: 2025-07-01. [Online]. Available: \url{https://wiki.haskell.org/Foreign_Function_Interface}
\BIBentrySTDinterwordspacing

\bibitem{python_ctypes}
\BIBentryALTinterwordspacing
{Python Software Foundation}. (2025) {ctypes — A foreign function library for Python}. Accessed: 2025-07-01. [Online]. Available: \url{https://docs.python.org/3/library/ctypes.html}
\BIBentrySTDinterwordspacing

\bibitem{cffi_docs}
\BIBentryALTinterwordspacing
{cffi developers}. (2025) {CFFI Documentation (Stable)}. Accessed: 2025-07-01. [Online]. Available: \url{https://cffi.readthedocs.io/en/stable/}
\BIBentrySTDinterwordspacing

\bibitem{python_extension_modules}
\BIBentryALTinterwordspacing
{Python Software Foundation}. (2025) {Extending Python with C or C++}. Accessed: 2025-07-01. [Online]. Available: \url{https://docs.python.org/3/extending/extending.html}
\BIBentrySTDinterwordspacing

\bibitem{swig_tool}
\BIBentryALTinterwordspacing
{SWIG Developers}. (2025) {SWIG: Simplified Wrapper and Interface Generator}. Accessed: 2025-07-01. [Online]. Available: \url{https://www.swig.org/}
\BIBentrySTDinterwordspacing

\bibitem{python_faulthandler}
\BIBentryALTinterwordspacing
{Python Software Foundation}. (2025) {faulthandler — Dump the Python traceback}. Accessed: 2025-07-01. [Online]. Available: \url{https://docs.python.org/3/library/faulthandler.html}
\BIBentrySTDinterwordspacing

\bibitem{FR19}
G.~Fraser and J.~M. Rojas, ``Software testing,'' in \emph{Handbook of Software Engineering}.\hskip 1em plus 0.5em minus 0.4em\relax Springer, 2019, pp. 123--192.

\bibitem{PLE+07}
C.~Pacheco, S.~K. Lahiri, M.~D. Ernst, and T.~Ball, ``Feedback-directed random test generation,'' in \emph{Proc.\ ICSE}.\hskip 1em plus 0.5em minus 0.4em\relax {IEEE} Comp. Soc., 2007, pp. 75--84.

\bibitem{Ton04}
P.~Tonella, ``Evolutionary testing of classes,'' in \emph{Proc.\ ISSTA}.\hskip 1em plus 0.5em minus 0.4em\relax {ACM}, 2004, pp. 119--128.

\bibitem{CGA18}
J.~Campos, Y.~Ge, N.~Albunian, G.~Fraser, M.~Eler, and A.~Arcuri, ``An empirical evaluation of evolutionary algorithms for unit test suite generation,'' \emph{Inf.\ Softw.\ Technol.}, vol. 104, pp. 207--235, 2018.

\bibitem{Panichella2018}
A.~Panichella, F.~M. Kifetew, and P.~Tonella, ``A large scale empirical comparison of state-of-the-art search-based test case generators,'' \emph{Inf.\ Softw.\ Technol.}, vol. 104, pp. 236--256, 2018.

\bibitem{DDS21}
S.~Dola, M.~B. Dwyer, and M.~L. Soffa, ``Distribution-aware testing of neural networks using generative models,'' in \emph{Proc.\ ICSE}.\hskip 1em plus 0.5em minus 0.4em\relax {IEEE}, 2021, pp. 226--237.

\bibitem{dakhel_effective_2024}
A.~M. Dakhel, A.~Nikanjam, V.~Majdinasab, F.~Khomh, and M.~C. Desmarais, ``Effective test generation using pre-trained {Large} {Language} {Models} and mutation testing,'' \emph{Inf.\ Softw.\ Technol.}, vol. 171, p. 107468, 2024.

\bibitem{yang_enhancing_2024}
C.~Yang, J.~Chen, B.~Lin, J.~Zhou, and Z.~Wang, ``Enhancing {LLM}-based test generation for hard-to-cover branches via program analysis,'' \emph{CoRR}, vol. abs/2404.04966, 2024.

\bibitem{pizzorno_coverup_2024}
J.~A. Pizzorno and E.~D. Berger, ``{CoverUp}: Effective high coverage test generation for python,'' \emph{Proc.\ {ACM} Softw.\ Eng.}, vol.~2, no. FSE, pp. 2897--2919, 2025.

\bibitem{ryan_code-aware_2024}
G.~Ryan, S.~Jain, M.~Shang, S.~Wang, X.~Ma, M.~K. Ramanathan, and B.~Ray, ``Code-aware prompting: A study of coverage-guided test generation in regression setting,'' \emph{Proc.\ {ACM} Softw.\ Eng.}, vol.~1, no. FSE, pp. 951--971, 2024.

\bibitem{yang_llm-enhanced_2025}
R.~Yang, X.~Xu, and R.~Wang, ``{LLM}-enhanced evolutionary test generation for untyped languages,'' \emph{Autom.\ Softw.\ Eng.}, vol.~32, no.~1, p.~20, 2025.

\bibitem{xiao_optimizing_2024}
D.~Xiao, Y.~Guo, Y.~Li, and L.~Chen, ``Optimizing search-based unit test generation with large language models: An empirical study,'' in \emph{Proc. Internetware}.\hskip 1em plus 0.5em minus 0.4em\relax {ACM}, 2024, pp. 71--80.

\bibitem{jain_testforge_2025}
K.~Jain and C.~L. Goues, ``{TestForge}: Feedback-driven, agentic test suite generation,'' \emph{CoRR}, vol. abs/2503.14713, 2025.

\bibitem{sapozhnikov_testspark_2024}
A.~Sapozhnikov, M.~Olsthoorn, A.~Panichella, V.~Kovalenko, and P.~Derakhshanfar, ``{TestSpark}: {IntelliJ} {IDEA}'s ultimate test generation companion,'' in \emph{Proc.\ ICSE Companion}.\hskip 1em plus 0.5em minus 0.4em\relax {ACM}, 2024, pp. 30--34.

\bibitem{abdullin_test_2025}
A.~Abdullin, P.~Derakhshanfar, and A.~Panichella, ``Test wars: A comparative study of {SBST}, symbolic execution, and {LLM}-based approaches to unit test generation,'' in \emph{Proc.\ ICST}.\hskip 1em plus 0.5em minus 0.4em\relax {IEEE}, 2025, pp. 221--232.

\bibitem{chen_chatunitest_2024}
Y.~Chen, Z.~Hu, C.~Zhi, J.~Han, S.~Deng, and J.~Yin, ``{ChatUniTest}: A framework for {LLM}-based test generation,'' in \emph{Proc.\ FSE Companion}.\hskip 1em plus 0.5em minus 0.4em\relax {ACM}, 2024, pp. 572--576.

\bibitem{ouedraogo_llms_2024}
W.~C. Ouedraogo, K.~Kabore, H.~Tian, Y.~Song, A.~Koyuncu, J.~Klein, D.~Lo, and T.~F. Bissyande, ``{LLMs} and prompting for unit test generation: A large-scale evaluation,'' in \emph{Proc.\ ASE}.\hskip 1em plus 0.5em minus 0.4em\relax {ACM}, 2024, pp. 2464--2465.

\bibitem{roychowdhury_static_2025}
S.~Roychowdhury, G.~Sridhara, A.~K. Raghavan, J.~Bose, S.~Mazumdar, H.~Singh, S.~B. Sugumaran, and R.~Britto, ``Static program analysis guided {LLM} based unit test generation,'' \emph{CoRR}, vol. abs/2503.05394, 2025.

\bibitem{schafer_empirical_2024}
M.~Schäfer, S.~Nadi, A.~Eghbali, and F.~Tip, ``An empirical evaluation of using large language models for automated unit test generation,'' \emph{{IEEE} Trans.\ Software Eng.}, vol.~50, no.~1, pp. 85--105, 2024.

\bibitem{PKT18b}
A.~Panichella, F.~M. Kifetew, and P.~Tonella, ``Automated test case generation as a many-objective optimisation problem with dynamic selection of the targets,'' \emph{{IEEE} Trans.\ Software Eng.}, vol.~44, no.~2, pp. 122--158, 2018.

\bibitem{dill_github}
\BIBentryALTinterwordspacing
{UQFoundation}. (2025) {dill}. Accessed: 2025-06-16. [Online]. Available: \url{https://github.com/uqfoundation/dill}
\BIBentrySTDinterwordspacing

\bibitem{McKerns2010}
\BIBentryALTinterwordspacing
M.~McKerns and M.~Aivazis, ``pathos: a framework for heterogeneous computing,'' 2010-. [Online]. Available: \url{https://uqfoundation.github.io/project/pathos}
\BIBentrySTDinterwordspacing

\bibitem{McKerns2011}
M.~M. McKerns, L.~Strand, T.~Sullivan, A.~Fang, and M.~A.~G. Aivazis, ``Building a framework for predictive science,'' \emph{CoRR}, vol. abs/1202.1056, 2011.

\bibitem{pathos_github}
\BIBentryALTinterwordspacing
{UQFoundation}. (2025) {pathos: multiprocess module}. Accessed: 2025-06-16. [Online]. Available: \url{https://github.com/uqfoundation/pathos}
\BIBentrySTDinterwordspacing

\bibitem{python_inspect_getsource}
\BIBentryALTinterwordspacing
{Python Software Foundation}. (2025) {inspect.getsource — Python Docs}. Accessed: 2025-07-15. [Online]. Available: \url{https://docs.python.org/3/library/inspect.html#inspect.getsource}
\BIBentrySTDinterwordspacing

\bibitem{cloc_tool}
\BIBentryALTinterwordspacing
{AlDanial}. (2025) {cloc: Count Lines of Code}. Accessed: 2025-07-09. [Online]. Available: \url{https://github.com/AlDanial/cloc}
\BIBentrySTDinterwordspacing

\bibitem{github_api_languages}
\BIBentryALTinterwordspacing
{GitHub}. (2025) {GitHub REST API: List repository languages}. Accessed: 2025-07-09. [Online]. Available: \url{https://docs.github.com/en/rest/repos/repos?apiVersion=2022-11-28#list-repository-languages}
\BIBentrySTDinterwordspacing

\bibitem{taskaya_reiz_2021}
B.~Taskaya, ``Reiz: Structural source code search,'' \emph{J. Open Source Softw.}, vol.~6, no.~62, p. 3296, 2021.

\bibitem{top_pypi_packages}
\BIBentryALTinterwordspacing
{Hugo van Kemenade}. (2025) {Top PyPI Packages}. Accessed: 2025-05-12. [Online]. Available: \url{https://hugovk.github.io/top-pypi-packages/}
\BIBentrySTDinterwordspacing

\bibitem{AB14}
A.~Arcuri and L.~C. Briand, ``A hitchhiker's guide to statistical tests for assessing randomized algorithms in software engineering,'' \emph{Softw.\ Test.\ Verification Reliab.}, vol.~24, no.~3, pp. 219--250, 2014.

\bibitem{mann_test_1947}
H.~B. Mann and D.~R. Whitney, ``On a test of whether one of two random variables is stochastically larger than the other,'' \emph{The Annals of Mathematical Statistics}, vol.~18, no.~1, pp. 50--60, 1947, publisher: Institute of Mathematical Statistics.

\bibitem{VD00}
A.~Vargha and H.~D. Delaney, ``A critique and improvement of the cl common language effect size statistics of mcgraw and wong,'' \emph{J.\ Educ.\ Behav.\ Stat.}, vol.~25, no.~2, pp. 101--132, 2000.

\bibitem{gruber_automatic_2024}
M.~Gruber, M.~F. Roslan, O.~Parry, F.~Scharnböck, P.~McMinn, and G.~Fraser, ``\BIBforeignlanguage{en}{Do automatc test generation tools generate flaky tests?}'' in \emph{\BIBforeignlanguage{en}{Proc.\ ICSE}}.\hskip 1em plus 0.5em minus 0.4em\relax {ACM}, 2024, pp. 1--12.

\bibitem{python_errors_exceptions}
\BIBentryALTinterwordspacing
{Python Software Foundation}. (2025) {Intermezzo: Errors and Exceptions}. Accessed: 2025-07-01. [Online]. Available: \url{https://docs.python.org/3/extending/extending.html#intermezzo-errors-and-exceptions}
\BIBentrySTDinterwordspacing

\bibitem{linux_man_signal_7}
\BIBentryALTinterwordspacing
{Michael Kerrisk}. (2025) {signal(7) - Linux manual page}. Accessed: 2025-07-18. [Online]. Available: \url{https://man7.org/linux/man-pages/man7/signal.7.html}
\BIBentrySTDinterwordspacing

\bibitem{rosler_reconstructing_2013}
J.~Rößler, A.~Zeller, G.~Fraser, C.~Zamfir, and G.~Candea, ``Reconstructing core dumps,'' in \emph{Proc.\ ICST}.\hskip 1em plus 0.5em minus 0.4em\relax {IEEE}, 2013, pp. 114--123.

\bibitem{soltani_guided_2017}
M.~Soltani, A.~Panichella, and A.~van Deursen, ``A guided genetic algorithm for automated crash reproduction,'' in \emph{Proc.\ ICSE}.\hskip 1em plus 0.5em minus 0.4em\relax {IEEE}, 2017, pp. 209--220.

\bibitem{derakhshanfar_botsing_2021}
P.~Derakhshanfar, X.~Devroey, A.~Panichella, A.~Zaidman, and A.~van Deursen, ``Botsing, a search-based crash reproduction framework for {Java},'' in \emph{Proc.\ ASE}, 2021, pp. 1278--1282.

\bibitem{bergel_beacon_2021}
A.~Bergel and I.~S. Munoz, ``Beacon: {Automated} {Test} {Generation} for {Stack}-{Trace} {Reproduction} using {Genetic} {Algorithms},'' in \emph{Proc.\ SBSE@ICSE}.\hskip 1em plus 0.5em minus 0.4em\relax {IEEE}, 2021.

\bibitem{arcuri_restful_2019}
A.~Arcuri, ``{RESTful} {API} automated test case generation with {EvoMaster},'' \emph{{ACM} Trans.\ Softw.\ Eng.\ Methodol.}, vol.~28, no.~1, pp. 3:1--3:37, 2019.

\bibitem{kechagia_effective_2019}
M.~Kechagia, X.~Devroey, A.~Panichella, G.~Gousios, and A.~van Deursen, ``Effective and efficient {API} misuse detection via exception propagation and search-based testing,'' in \emph{Proc.\ ISSTA}.\hskip 1em plus 0.5em minus 0.4em\relax {ACM}, 2019, pp. 192--203.

\end{thebibliography}

\end{document}